\documentclass{aa}  
\usepackage{color} 
\usepackage[squaren, Gray, cdot]{SIunits}
\usepackage{graphicx}
\usepackage{txfonts}
\usepackage{natbib}
\bibpunct{(}{)}{;}{a}{}{,} % to follow the A&A style
\usepackage{amsmath}
\newcounter{eqn}

\def \cc    {\ifmmode{\,{\rm cm}^{-3}}\else{$\,{\rm cm}^{-3}$}\fi}
\def \ccs   {\ifmmode{\,{\rm cm}^{-3} {\rm s}^{-1}}\else{$\,{\rm cm}^{-3} {\rm s}^{-1}$}\fi}
\def \cq    {\ifmmode{\,{\rm cm}^{-2}}\else{$\,{\rm cm}^{-2}$}\fi}
\def \mic   {\ifmmode{\,\mu{\rm m}}\else{$\mu$m}\fi}
\def \eccs  {\ifmmode{\,{\rm erg}\,{\rm cm}^{-3} {\rm s}^{-1}}\else{$\,{\rm erg}\,{\rm cm}^{-3} {\rm s}^{-1}$}\fi}
\def \ecc   {\ifmmode{\,{\rm erg}\,{\rm cm}^{-3}}\else{$\,{\rm erg}\,{\rm cm}^{-3}$}\fi}
\def \ecm   {\ifmmode{\,{\rm erg}\,{\rm cm}^{-1}}\else{$\,{\rm erg}\,{\rm cm}^{-1}$}\fi}
\def \ecqs  {\ifmmode{\,{\rm erg}\,{\rm cm}^{-2}\,{\rm s}^{-1}\,{\rm sr}^{-1}}\else{$\,{\rm erg}\,{\rm cm}^{-2}\,{\rm s}^{-1}\,{\rm sr}^{-1}$}\fi}
\def \deg   {\ifmmode{^{\circ}}\else{$^{\circ}$}\fi} 
\def \pc    {\ifmmode{\,{\rm pc}}\else{$\,{\rm pc}$}\fi} 
\def \kms   {\ifmmode{\,{\rm km}\,{\rm s}^{-1}}\else{km s$^{-1}$}\fi} 
\def \kmspc {\ifmmode{\,{\rm km}\,{\rm s}^{-1}\,{\rm pc}^{-1}}\else{km s$^{-1}$ pc$^{-1}$}\fi} 
\def \MJysr {\ifmmode{\,{\rm MJy\,sr}^{-1}}\else{$\,{\rm MJy\,sr}^{-1}$}\fi} 
\def \Kkms  {\ifmmode{\,{\rm K\,km\,s}^{-1}}\else{$\,{\rm K\,km\,s}^{-1}$}\fi}
\def \dens{\ifmmode{n_{\rm H}}\else{$n_{\rm H}$}\fi}
\def \epso{\ifmmode{\overline{\varepsilon}}\else{$\overline{\varepsilon}$}\fi}
\def \utM{\ifmmode{u_{\theta,{\rm M}}}\else{$u_{\theta,{\rm M}}$}\fi}
\def \urM{\ifmmode{u_{r,{\rm M}}}\else{$u_{r,{\rm M}}$}\fi}
\def \zetaHH{\ifmmode{\zeta_{\HH}}\else{$\zeta_{\HH}$}\fi}
\def \zetaH {\ifmmode{\zeta_{\rm H}}\else{$\zeta_{\rm H}$}\fi}
\def \HH{\ifmmode{\rm H_2}\else{${\rm H}_2$}\fi}
\def \Cp{\ifmmode{\rm C^+}\else{${\rm C}^+$}\fi} 
\def \Sp{\ifmmode{\rm S^+}\else{${\rm S}^+$}\fi} 
\makeatletter
\newcommand{\putindeepbox}[2][0.7\baselineskip]{{%
    \setbox0=\hbox{#2}%
    \setbox0=\vbox{\noindent\hsize=\wd0\unhbox0}
    \@tempdima=\dp0
    \advance\@tempdima by \ht0
    \advance\@tempdima by -#1\relax
    \dp0=\@tempdima
    \ht0=#1\relax
    \box0
}}
\makeatother
\begin{document} 

\defcitealias{valdivia2016}{Paper~I}
\defcitealias{myers2015}{MML15}
%\addbibresource{biblio_val.bib}

   \title{Origin of CH$^+$ in diffuse molecular clouds}

  \subtitle{Warm H$_2$ and ion-neutral drift}

   \author{Valeska Valdivia\inst{\ref{inst1},\ref{inst2}}
        \and
        Benjamin Godard\inst{\ref{inst2},\ref{inst3}}
        \and
        Patrick Hennebelle\inst{\ref{inst1},\ref{inst2}}
        \and
        Maryvonne Gerin\inst{\ref{inst2},\ref{inst3}}
        \and
        Pierre Lesaffre\inst{\ref{inst2},\ref{inst3}}
        \and
        Jacques Le Bourlot\inst{\ref{inst2},\ref{inst4}}
          }

        \institute{Laboratoire AIM, Paris-Saclay, CEA/IRFU/SAp - CNRS - Universit\'e Paris Diderot, 91191 Gif-sur-Yvette Cedex, France\\
        \email{valeska.valdivia@lra.ens.fr} \label{inst1}
        \and 
        Laboratoire de radioastronomie, LERMA, Observatoire de Paris, \'Ecole Normale Sup\'erieure (UMR 8112 CNRS), 24 rue Lhomond, 75231 Paris Cedex 05, France
        \label{inst2}
        \and 
        Sorbonne Universit\'es, UPMC Univ Paris 06, UMR 8112, LERMA, 75005 Paris, France \label{inst3}
        \and 
        Universit\'e Paris Diderot, Sorbonne Paris Cit\'e, 75013 Paris, France \label{inst4}
        }

   \date{Received Month dd, yyyy; accepted Month dd, yyyy}

% \abstract{}{}{}{}{} 
% 5 {} token are mandatory
 
  \abstract
   {
%The large abundances of CH$^+$ observed in the local diffuse interstellar gas have been a longstanding problem. The main formation path is the ion-neutral reaction between C$^+$ and H$_2$, which is is highly endothermic. This reaction requires either warm reactants or non-thermal processes to occur efficiently. 
Molecular clouds are known to be magnetised and to display a turbulent and complex structure where warm and cold phases are interwoven.
The turbulent motions within molecular clouds transport molecules, and the presence of magnetic fields induces a relative velocity between neutrals and ions known as the ion-neutral drift ($v_\mathrm{d}$). These effects all together can influence the chemical evolution of the clouds.}
   {This paper  assesses the roles of two physical phenomena which have previously been invoked to boost the production of CH$^+$ under realistic physical conditions: the presence of warm H$_2$ and the increased formation rate due to the ion-neutral drift.}
   {We performed ideal magnetohydrodynamical (MHD) simulations that include the heating and cooling of the multiphase interstellar medium (ISM), and where we treat dynamically the formation of the \HH\ molecule. In a post-processing step we compute the abundances of species at chemical equilibrium using a solver that we developed. The solver uses the physical conditions of the gas as input parameters, and can also prescribe the \HH\ fraction if needed. We validate our approach by showing that  the H$_2$ molecule generally has a much longer chemical evolution timescale compared to the other species.}
   {
%We present the ion-neutral drift velocity distributions, as well as the distribution of CH$^+$ as a function of the gas temperature, the fraction of H$_2$, and the drift velocity. We also compare the total column densities of CH$^+$, obtained in the simulations, to available observations.
%We present the ion-neutral drift velocity distributions, as well as the distribution of CH$^+$
        We show that CH$^+$ is efficiently formed at the edge of clumps, in regions where the H$_2$ fraction is low ($0.3-30\%$) but nevertheless higher than its equilibrium value, and where the gas temperature is high ($\gtrsim 300~\mathrm{K}$). 
%These conditions are fulfilled at the edge of clumps, where H$_2$ is found to be out-of-equilibrium, and consequently the abundance of CH$^+$ is increased by several orders of magnitude compared to the case where H$_2$ is at equilibrium.  
        We show that warm and out of equilibrium H$_2$ increases the integrated column densities of CH$^+$ by one order of magnitude up to values still $\sim3-10$ times lower than those observed in the diffuse ISM.
%but the abundances are still underpredicted by a factor of $\sim3-10$ compared to observations. 
%       We balance the Lorentz force with the ion-neutral drag to estimate the ion-drift velocities from our ideal MHD simulations. We find that the ion-neutral drift velocity distribution peaks around $\sim 0.04~\mathrm{km\ s^{-1}}$, and high drift velocities are rare, leading to a negligible statistical impact on the abundances of CH$^+$. Our multiphase simulations reduce the spread in $v_\mathrm{d}$ compared to Myers et al, and our self-consistent treatment of the ionisation leads to much reduced $v_\mathrm{d}$ compared to their work. Nevertheless, this distribution depends on the numerical resolution, which is insufficient to obtain a converged velocity distribution, and higher resolution simulations could show a more significative influence.
        We balance the Lorentz force with the ion-neutral drag to estimate the ion-drift velocities from our ideal MHD simulations. We find that the ion-neutral drift velocity distribution peaks around $\sim 0.04~\mathrm{km\ s^{-1}}$, and that high drift velocities are too rare to have a significant statistical impact on the abundances of CH$^+$. 
%
%, leading to a negligible statistical impact on the abundances of CH$^+$. 

Compared to previous works, our multiphase simulations reduce the spread in $v_\mathrm{d}$, and our self-consistent treatment of the ionisation leads to much reduced $v_\mathrm{d}$. Nevertheless, our resolution study shows that this velocity distribution is not converged:
%, with a larger impact of ion-neutral drift on CH$^+$ at higher resolution. \textbf{ ... converged: 
the ion-neutral drift has
a higher impact  on CH$^+$ at higher resolution.  On the other hand, our ideal MHD simulations do not include ambipolar diffusion, which would yield lower drift velocities.
%
%as well as the distribution of CH$^+$. 
}
{Within these limitations, we conclude that warm H$_2$ is a key ingredient in the efficient formation of CH$^+$ and that the ambipolar diffusion has very little influence on the abundance of CH$^+$, mainly due to the small drift velocities obtained. However, we point out that small-scale processes and other non-thermal processes not included in our MHD simulation may be of crucial importance, and higher resolution studies with better controlled dissipation processes are needed.}

   \keywords{CH$^+$--
                H$_2$ --
                molecular clouds --
                ISM
                }

   \maketitle
%
%-------------------------------------------------------------------

\section{Introduction}

It is widely accepted that molecular clouds are multiphase entities where the dense and cold gas, called the cold neutral medium (CNM), is mixed to a more diffuse and warm phase, known as the warm neutral medium (WNM). Interstellar turbulence not only leads to large density fluctuations \citep{falgarone1995}, but it also mixes up these two phases. Using 1D simulations \citet{lesaffre2007} have shown that the turbulent diffusion can transport molecules into warmer regions by spreading out the transition layers between the CNM and the WNM. More recently, in \citet[][hereafter Paper I]{valdivia2016}, we showed a similar effect in 3D magnetohydrodynamical (MHD) simulations of turbulent molecular clouds. In particular we showed that molecular hydrogen (H$_2$) can be formed at intermediate densities in transient structures and subsequently carried toward warmer regions by the turbulent motions within the gas. This warm H$_2$ is in good agreement with observations of excited populations of H$_2$ in translucent molecular clouds \citep{rachford2002, gry2002, lacour2005}, and it is reasonably successful at explaining these populations as collisionally excited states. 

%=======================================================
%A possible related problem is the formation of the methylidine cation CH$+$ in the interstellar medium (ISM), which requires molecular hydrogen to be formed efficiently. CH$^+$ was one of the first molecules observed in space \citep{douglas1941}, and it has been detected in a wide variety of environments going from the interstellar medium in the Milky Way in visible wavelengths in the local ISM \citep{crane1995, gredel1997, weselak2008}, and in infrared wavelengths deeper in the Galactic disc \citep{falgarone2005, falgarone2010, godard2012},  to external galaxies \citep{rangwala2011, vanderwerf2010, ritchey2015, spinoglio2012}. But its ubiquity and high abundance in the diffuse environments in the ISM has been a puzzling problem for more than 70 years.  
%
A possible related problem is the formation of the methylidine cation CH$^+$ in the interstellar medium (ISM), which requires molecular hydrogen to be formed efficiently. CH$^+$ was one of the first molecules observed in space  \citep{douglas1941}, and it has been detected in a wide variety of environments, but its ubiquity and high abundance in the diffuse environments in the ISM has been a puzzling problem for more than 70 years.   
In the Milky Way, $^{12}\mathrm{CH^+}$ has been detected in visible wavelengths in the local ISM \citep{crane1995, gredel1997, weselak2008}, while the $^{13}\mathrm{CH}^+$ isotopologue has been detected in infrared wavelengths deeper in the Galactic disc \citep{falgarone2005, falgarone2010, godard2012}. This molecule has even been observed in the interstellar medium of external galaxies \citep{rangwala2011, vanderwerf2010, ritchey2015, spinoglio2012}. 

CH$^+$ is easily destroyed by reactions with electrons and hydrogen atoms, but also by reactions with H$_2$ molecules.  Under these conditions, the only reaction efficient enough to counterbalance the fast destruction is 
%The main formation path,
\begin{equation}
\mathrm{C^+ + H_2 \rightarrow CH^+ + H \qquad (\Delta E/k = -4300~K}).
\end{equation}

%%% A more accurate value of the endothermicity is 4300K as reported in Agundez et al. 2010. See also ARAA

\noindent As this reaction is highly endothermic \citep{agundez2010}, it has been proposed that the observed abundances might be related to warm layers of gas resulting, for instance, from turbulent mixing or turbulent dissipation processes \citep{crane1995, gredel1997, lambert1986}. 

%BG ==> 
From the theoretical point of view, the attemps to explain the large amounts of CH$^+$ in the diffuse ISM include shock heated gas \citep{draine1986, pineaudesforets1986, flower1998,lesaffre2013}, turbulent dissipation in vortices \citep{joulain1998, godard2009, godard2014, falgarone2010, myers2015}, and the turbulent mixing between the CNM and the WNM \citep{lesaffre2007}.  Several of these models have proven
successful in reproducing not only the abundances of CH$^+$, but also those of many other species (e.g. SH$^+$, CO, HCO$^+$) and their correlations observed in the diffuse medium under the constraint of the mechanical energy injected at large scale. However, they did so assuming idealised 1D steady-state structures and adopting either a single type of structure along the line of sight or very simplistic velocity, density, or magnetic field distributions with no link to the large sale dynamics of the gas.

%Among the weaknesses of most models we can cite the lack of realistic 3-dimensional structures, and the steady-state treatment of the chemistry. Several attemps to couple the chemistry and the 3-dimensional dynamics have been performed by other groups \citep{glover2010, grassi2014, seifried2016, ziegler2016}. 

%The recent work of \citet{myers2015} succeeded at reproducing the observed CH$^+$ abundances for a turbulent molecular cloud with typical physical conditions for the ISM. Using ideal magnetohydrodynamical (MHD) simulations, they showed that when the ion-neutral drift reaches $3-4~\mathrm{km\ s^{-1}}$, the formation of CH$^+$ is dominated by the drift between C$^+$ and H$_2$, and that the CH$^+$ column densities are comparable to the observed ones,  but whether the assumptions about the chemical and physical state of the gas are realistic is a matter that can be debated. Particularly the simplifying assumptions, such as the constant values of the fraction of molecular gas and the ionisation fraction, can eventually influence the predicted abundance of CH$^+$.

In their recent work,  \citet{myers2015} were able to reproduce the observed CH$^+$ abundances for a turbulent molecular cloud with typical physical conditions for the ISM. Using ideal magnetohydrodynamical (MHD) simulations, they showed that almost all of the CH$^+$ molecules are produced in regions where the ion-neutral drift velocities are high ($v_\mathrm{d} \gtrsim 3-4~\mathrm{km\ s^{-1}}$). CH$^+$ production is controlled by drift velocities higher than $3~\mathrm{km\ s^{-1}}$ which, although rare, are present in a sufficient number to be statistically significant and to produce CH$^+$ column densities comparable to observations. However, there are several caveats related to the simplifying assumptions that might be influencing their results. More specifically, the ionisation fraction is assumed to be constant and equal to the ionisation fraction of the dense phase, which might overestimate the drift velocities in the diffuse phase (where most of the CH$^+$ is produced). Another incorrect physical assumption is that the H$_2$ molecular fraction is assumed to be $10\%$ throughout the entire domain. Taken all together, these assumptions likely lead to an artificially high CH$^+$ abundance.

The goal of our paper is to study the formation of CH$^+$  on a large scale in diffuse molecular clouds. We use a dynamically calculated abundance of molecular hydrogen (out of equilibrium H$_2$) and include a more detailed description of the microphysical processes. We  compare our data with available observations and with the previous theoretical work of \citet[][hereafter MML15]{myers2015}. First, we explore the role of the turbulent mixing, which transports molecular hydrogen to the warm gas, in a realistic 3D structure by assessing what role is played by the warm reactants. Second, we address the question of whether the approximated treatment of the ambipolar diffusion in the ideal MHD limit can explain the observed abundances of CH$^+$ under realistic physical conditions. 

%we want to address the following questions: firstly, which is the role played by the turbulent mixing in the ISM, or in other words, to asses the role played by the warm reactants. Secondly, can the ambipolar diffusion explain the observed abudances of CH$^+$ under realistic physical conditions. 

The paper is structured as follows. In Sect.~\ref{methodology} we describe the numerical simulation and the chemical solver used to post-process it. We also describe how we estimate the ion-neutral slip velocity, and we analyse the validity of our approach.
% [H2 CHEM ON-THE-FLY, REST IN POSTPROC].
 In Sect.~\ref{results} we present our results on the importance of warm reactants, and the importance of the ion-neutral drift, on the abundance of CH$^+$. We conclude the paper in Sect.~\ref{conclusions}. We give a detailed description of our chemical solver and a detailed study of the chemical timescales for each individual species in the Appendix.

%\\
%Methodology\\
%Results\\
%Conclusions

%Additionally, the ion-neutral drift has been invoked as a manner to increase the production of CH$+$ without a further increasie in the gas temperature .
%A particularly 
%One of the relevant molecules in the study of this warm chemistry is the CH$^+$ 

%==> WRITE INTRODUCTION:

%- why CH+
%
%-ref: OBSERVATIONAL
%
%-ref: THEORETICAL 

%The formation of CH$^+$ is one of the key steps in the formation of more complex hydrocarbons in diffuse interstellar clouds. This was one of the first molecular species observed in the Interstellar medium (ISM) \citep{douglas1941}, and since then its ubiquity and high abundance in the ISM has been a puzzling problem. \citet{gredel1993} highlighted the requirement of "hot" reactants to form CH$^+$, arguing that the "hot" population of H$_2$ would be close to the amount needed to produce the observed CH$^+$ in translucent clouds, and that turbulence would play a key role. In the same direction, \citet{crane1995} presented high resolution observations of interstellar CH and CH$^+$, being able to resolve all the components of the spectra. They found that $b$ Doppler broadening parameters exceeded the expected thermal broadening. Both studies suggest that the observed abundances of CH$^+$ are not produced in shocks. \citet{godard2014} presented a chemical model driven by turbulence dissipation (TDR model), that proposes an alternative to traditional formation pathways (UV and CR driven chemistry).

%--------------------------------------------------------------------
\section{Methodology} \label{methodology}

\subsection{MHD simulation}

\begin{figure*}
\centering
\includegraphics[width=0.85\textwidth]{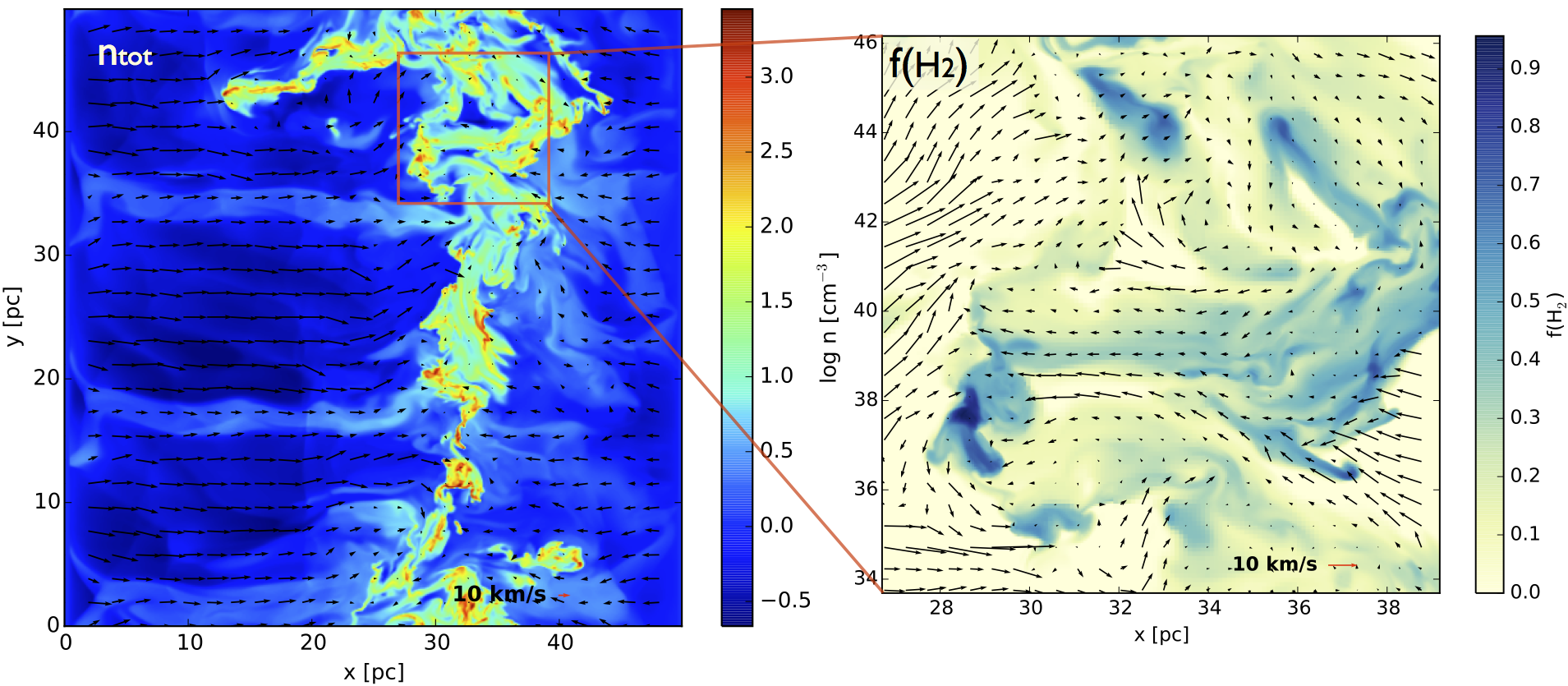}
\caption{Slice showing the total number density and the H$_2$ fraction. The arrows depict the velocity field projected onto the $x-y$ plane.}
\label{slice_sim_fig}
\end{figure*}

The numerical simulation used in this work is an ideal\footnote{It should be remembered  that non-ideal MHD simulations are out of reach of current computational capabilities, requiring complicated implicit or semi-implicit schemes, or extremely small time steps.} magnetohydrodynamical (MHD) multiphase simulation of a realistic molecular cloud using the RAMSES code \citep{teyssier2002}; it is  fully described in the Appendix of \citetalias{valdivia2016}.  

The molecular cloud is formed by colliding streams of atomic gas and the general set-up is 
very similar to \citet{valdivia2014} \citep[see also][]{audit2005, hennebelle2008}. The simulation box is a cube 
of side $L=50~\mathrm{pc}$ filled with atomic gas of density $n=1~\mathrm{cm^{-3}}$ and 
temperature $T=8000~\mathrm{K}$, embedded in the interstellar radiation field (ISRF) of strength $G_0=1.7$ in Habing units \citep{habing1968}, where two converging streams of WNM are injected from 
the $x$ boundaries with a slightly turbulent stationary velocity field of module $V_0 = 15~\mathrm{km\ s^{-1}}$.
% The corresponding injection of mechanical energy at the integral scale is 
%$\sim 2\times 10^{-25}$ \eccs\ [CHECK VALESKA], in agreement with the mean value of the 
%kinetic energy transfer rate deduced from the observed velocity dispersions of molecular 
%clouds seen in CO \citep{hennebelle2012}. The magnetic field is initially aligned with 
%the $x$ direction and it has an initial strength of $2.5~\micro G$.
 The corresponding injection of mechanical energy at the integral scale is 
$\dot{E}_\mathrm{in}\sim 3\times10^{33}~\mathrm{erg\ s^{-1}}$, or equivalently 
${\overline{\varepsilon}}_\mathrm{in}\sim 10^{-25}~\mathrm{erg\ cm^{-3} s^{-1}}$, 
in agreement with the mean value of the 
kinetic energy transfer rate deduced from the observed velocity dispersions of molecular 
clouds seen in CO \citep{hennebelle2012}. The magnetic field is initially aligned with 
the $x$ direction and it has an initial strength of $2.5~\micro G$.

The simulation uses an adaptive mesh refinement (AMR) technique. The minimum and maximum 
resolution levels are $\ell_{min} = 9$ and $\ell_{max} = 11$, reaching an effective numerical 
resolution of $2048^3$ or, equivalently, a spatial resolution of some $0.025~\mathrm{pc}$. 
The refinement criterion is density, with density thresholds at $n_\mathrm{thresh} = 
50~\mathrm{cm^{-3}}$, and $n_\mathrm{thresh} = 100~\mathrm{cm^{-3}}$. The time step of the
simulation is variable and smaller than $\sim 180$ yr, i.e. small enough to follow the propagation of dynamical 
perturbations in the CNM and WNM ($\sim 10^4$ yr). % and at the interfaces ($\sim 10^3$ yr).

The simulation follows the formation and destruction of H$_2$ and its thermal feedback. 
The effects of shielding from ultraviolet radiation by dust particles and \HH\ absorption
(self-shielding) have both been included for the computation of the \HH\ photodissociation rate 
using our tree-based method, detailed in \citet{valdivia2014} and in \citetalias{valdivia2016}. 
The simulation produces turbulent and highly structured molecular clouds that exhibit a 
wide range of physical conditions (in density, temperature, and shielding parameters).
Important fractions of out-of-equilibrium H$_2$ are found in the diffuse and warm phase of 
the gas as a consequence of the turbulent mixing and the shielding against photodissociation 
provided by the multiphase structure. Molecules of  \HH\  initially formed in transient dense
regions are spread by turbulence into the interclump medium where they can survive thanks to 
the shielding provided by the cloud structure. The presence of \HH\ in a warm environment
increases its excitation (\citetalias{valdivia2016}) and leads to column densities of \HH\
rotational levels comparable to those observed with the Copernicus and FUSE Telescopes
(e.g. \citealt{spitzer1974, frisch1980, frischjura1980, lambert1986, gry2002, lacour2005}). 

The snapshot used in this work corresponds to an evolution time of $15~\mathrm{Myr}$ in the simulation.
Figure~\ref{slice_sim_fig} shows the local number density in a cut through the middle plane, as well as a 
small region showing the local H$_2$ fraction. This figure depicts the influence of dynamics on 
the transport of H$_2$ molecules.

\subsection{Method description}

To understand the role of warm and out of equilibrium H$_2$ and of the multiphase structure 
of molecular clouds on the chemistry of the interstellar medium, we perform a post-treatment
of the numerical simulation described above. The data are extracted with the python module 
PyMSES \citep{labadens2012}. Once extracted, the chemical composition of every cell of the 
simulation is computed assuming chemical equilibrium for all species except for H and \HH.

The chemical solver used for the post-treatment is taken from the Meudon PDR (Photon 
Dominated Regions) code\footnote{Version 1.5.2 available at \url{http://ism.obspm.fr}.} 
(e.g. \citealt{lepetit2006,bron2014}), stripped of surface reaction processes and  
detailed treatments of grain physics and radiative transfer, and modified to allow the 
abundance of \HH\ to be fixed beforehand. For each cell, the solver takes as input the
local properties of the gas, i.e. the total hydrogen density $\dens$, the kinetic
temperature $T_K$, the external UV radiation field $\chi$, the visual extinction $A_V$, 
the ion-neutral velocity drift $v_\mathrm{d}$, the abundance of \HH\ $n(\HH)$, and 
the shielding of \HH\ from UV photons: $f_{\rm sh,\,\,\HH} = \langle 
\mathrm{e} ^{-\tau _{d, 1000}} f_\mathrm{shield} (\mathcal{N}_\mathrm{H_2})\rangle$
(\citetalias{valdivia2016}). 
%and $f_{\rm sh,\,\,CO} = \langle \mathrm{e} ^{-\tau _{d, 1000}} 
%f_\mathrm{shield} (\mathcal{N}_\mathrm{H_2},\mathcal{N}_\mathrm{CO})\rangle$. 
 The MHD simulation is ideal and hence the ion-neutral drift  also has to be estimated by a
 post-processing treatment (see Sect.~\ref{indrift} below). In output, 
the solver returns the at-equilibrium abundances of the 147 variable species included in 
the chemical network. Detailed descriptions and tests of the solver and of the method 
used to fix the value of $n(\HH)$ are presented in Appendix \ref{ap_chemsolv}. 

It is important to note that in this work we only consider the shielding of CO by dust,
and neglect both the self-shielding and the shielding induced by H$_2$ line absorption.
%component for CO, which is shielded by
%dust, and by H$_2$.

\subsection{Timescales and transient chemistry} \label{sect-timescale}

Computing the composition of the gas at chemical equilibrium with a fixed abundance of 
\HH\ is valid only if the timescales required to reach the equilibrium are smaller than 
that of \HH, but also smaller than the typical timescales of variation of dynamical 
quantities (\dens, $v$, $T_K$). To check these assumptions, the chemical timescales 
have been estimated in Appendix \ref{ap_chemsolv} as functions of the physical properties 
of the gas, in particular the density and temperature. These timescales (see 
Fig. \ref{Fig-tscale}) are found to vary between $10^2$ and $10^6$ yr depending on the 
species considered and the gas density. CH$^+$, for instance, reaches its equilibrium 
abundance in $\sim 2 \times 10^6 / \dens$ yr, regardless of the gas temperature.

A comparison with the results obtained for \HH\ shows that molecular hydrogen has 
an evolution time longer than that of any other species over almost the whole range 
of physical conditions spanned in the simulation. This result not only proves the 
existence of an equilibrium solution when the abundance of \HH\ is fixed, it also 
indicates that this solution is a coherent description of the chemical composition
of the gas over time $\geqslant 4 \times 10^6 / \dens$ yr for $\dens \leqslant 100$ 
\cc\ and $\geqslant 4 \times 10^4 (\dens/100)^{-0.23}$ yr for $\dens \geqslant 100$ 
\cc.

The comparison with dynamical timescales is, unfortunately, less satisfactory. With
the spatial resolution of the simulation, perturbations of physical conditions in
the WNM and CNM propagates at sound speed  over timescales of $\sim10^4$ yr. 
%At the interface between the two phases, the typical timescales
%of variation of the physical conditions can even be shorter ($\sim 10^3$ yr or less).
Therefore, while computing the chemistry at equilibrium may be marginally valid for
high-density gas ($\dens \geqslant 100$ \cc), it is probably not appropriate for cells
at lower density. In this case, an accurate treatment would require  following the 
time-dependent evolution of the entire chemistry, which is still out of reach of any
numerical simulation.

While the problem seems hopeless, we note that the chemical timescales presented in Appendix 
\ref{ap-timescales} have been calculated assuming that only \HH\ is out of equilibrium in 
the molecular clouds. It is probable, however, that only a few other species evolve, 
like \HH, over a long period, and need to be treated on the fly in the simulation. If so, 
computing the equilibrium abundances of all species except for a few critical ones would 
be a correct and very efficient method. Therefore, although our approach is not without fault, 
we argue that it gives a first estimation of the chemical state towards which the gas tends. 
It also provides a framework for future developments where other species could be considered 
out of equilibrium depending on the dynamical timescales, the properties of the gas, and the 
chemical history. All these developments are underway.

Another limitation of computing the chemistry at equilibrium 
is that it neglects any transient chemical event which can only be accounted for with time 
averaged abundances. However, we argue that taking into account such a transient chemical event within each fluid cell is not necessarily correct as long as the dynamics is not resolved over the same temporal scales.

\subsection{Ion-neutral drift} \label{indrift}

The ion-neutral drift, or ambipolar diffusion, can help to increase the rates of highly endothermic reactions which scale as $\exp({-\Delta E/kT})$, where $\Delta E$ is the endothermicity. The relative motion between neutrals and ions increases the effective velocity dispersion, which results in the increase of the effective temperature at which the reaction occurs \citep{draine1980, draine1983, flower1985}:
\begin{equation}
T_\mathrm{eff} = \frac{m_iT_n + m_n T_i}{m_i + m_n} + \Delta T, %\approx T + \frac{\mu}{3k} v_\mathrm{d}^2
\label{eq_Teff}
\end{equation}
 \noindent where $m_i$ ($m_n$) and $T_i$ ($T_n$)  are the mass and temperature of ions (neutrals), and $\Delta T$ is
\begin{equation}
\Delta T =  \frac{\mu}{3k} v_\mathrm{d}^2.
\label{deltaT}
\end{equation}
\noindent  In this expression $v_\mathrm{d}$ is the ion-neutral drift velocity, $\mu$ is the reduced mass of the reaction, and $k$ is the Boltzmann constant.
When ions and neutrals have  the same temperature (equal to the gas temperature, $T$), the effective temperature can be approximated as $T_\mathrm{eff} \approx T + \Delta T$. But when $T < \Delta T \approx \beta$, where $\beta = \Delta E/k$ is the endothermicity expressed in kelvins, this approximation overestimates the reaction rate \citep{pineaudesforets1986}, and the approximation is no longer valid. A more accurate result is obtained by taking the ion-neutral reaction rate to be proportional to $\exp \left(-\max\left\{\beta/T_\mathrm{eff}, (\beta - 3\Delta T)/T\right\}\right) $.\\

% In Appendix \ref{ap_gpdf} we show that for the physical conditions of our simulation this approximation is accurate enough.

%Given that in the ideal MHD approximation ions and neutrals are treated as a single fluid, their relative velocity can be estimated from the Maxwell's equation of induction and from the ion equation of motion as:

Although our simulation is in the ideal MHD approximation, we estimate the ion-neutral drift by neglecting the ion inertia in the ion equation of motion. The resulting balance between Lorentz forces and the ion-neutral drag force yields \citep{shang2002,glassgold2005}

\begin{equation}
\frac{\left( \nabla \times \vec{B}\right) \times \vec{B}}{4\pi} = \gamma \rho_n \rho_i v_\mathrm{d} = \sum_{jk} n_j n_k \mu_{jk} v_{\mathrm{d}, jk} K_{jk},
\label{eq_vAD}
\end{equation}

%\begin{equation}
%v_\mathrm{d} \approx \frac{\left( \nabla \times \vec{B}\right) \times \vec{B}}{4\pi \rho_{n} \rho_{i} \gamma}, %= \frac{\left( \nabla \times \vec{B}\right) \times \vec{B}}{4\pi n_{n} n_\mathrm{e} m_\mathrm{H}^{2} \gamma}
%\label{eq_vAD}
%\end{equation}
% 
\noindent where $\vec{B}$ is the magnetic field, $\rho_n$ and $\rho_i$ are respectively the densities of neutrals and ions, $\gamma$ is the coefficient for ambipolar diffusion,
%
%The ambipolar diffusion coefficient, $\gamma$, can be estimated from the expression for the drag force $f_\mathrm{D}$ \citep{shang2002,glassgold2005}:
%%For estimating the ambipolar diffusion coefficient, $\gamma$, we follow \citet{shang2002} and \citet{glassgold2005}. The expression for the drag force $f_D$ :
%%
%\begin{equation}
%f_\mathrm{D} = \gamma \rho_n \rho_i v_\mathrm{d} = \sum_{jk} n_j n_k \mu_{jk} v_{\mathrm{d}, jk} K_{jk}.
%\end{equation}
%%
%\noindent 
$n_j$ and $n_k$ are the number densities of the ionic and neutral species, $v_{\mathrm{d}, jk} = |v_\mathrm{j} - v_\mathrm{k}|$ is the specific drift velocity for species $j$ and $k$, $\mu_{jk}$ is the reduced mass, and $K_{jk}$ is the momentum transfer rate coefficient. Assuming the same drift velocity for all the species, $v_\mathrm{d}$ can be approximated as follows:
\begin{equation}
%\gamma  = \frac{ \sum_{jk} { n_j  n_k  \mu_{jk}  K_{jk} }}{\sum_{jk}{ n_j   m_j   n_k m_k }} \label{eq:ga}
v_\mathrm{d} \approx \frac{\left( \nabla \times \vec{B}\right) \times \vec{B}} {4\pi \sum_{jk} n_j n_k \mu_{jk} K_{jk}}
\label{eq:driftvel}
\end{equation}  
%
%%%%%%%%  FIG RESULTS %%%%%%%%%
\begin{figure*}[ht]
\centering
\includegraphics[width=6cm]{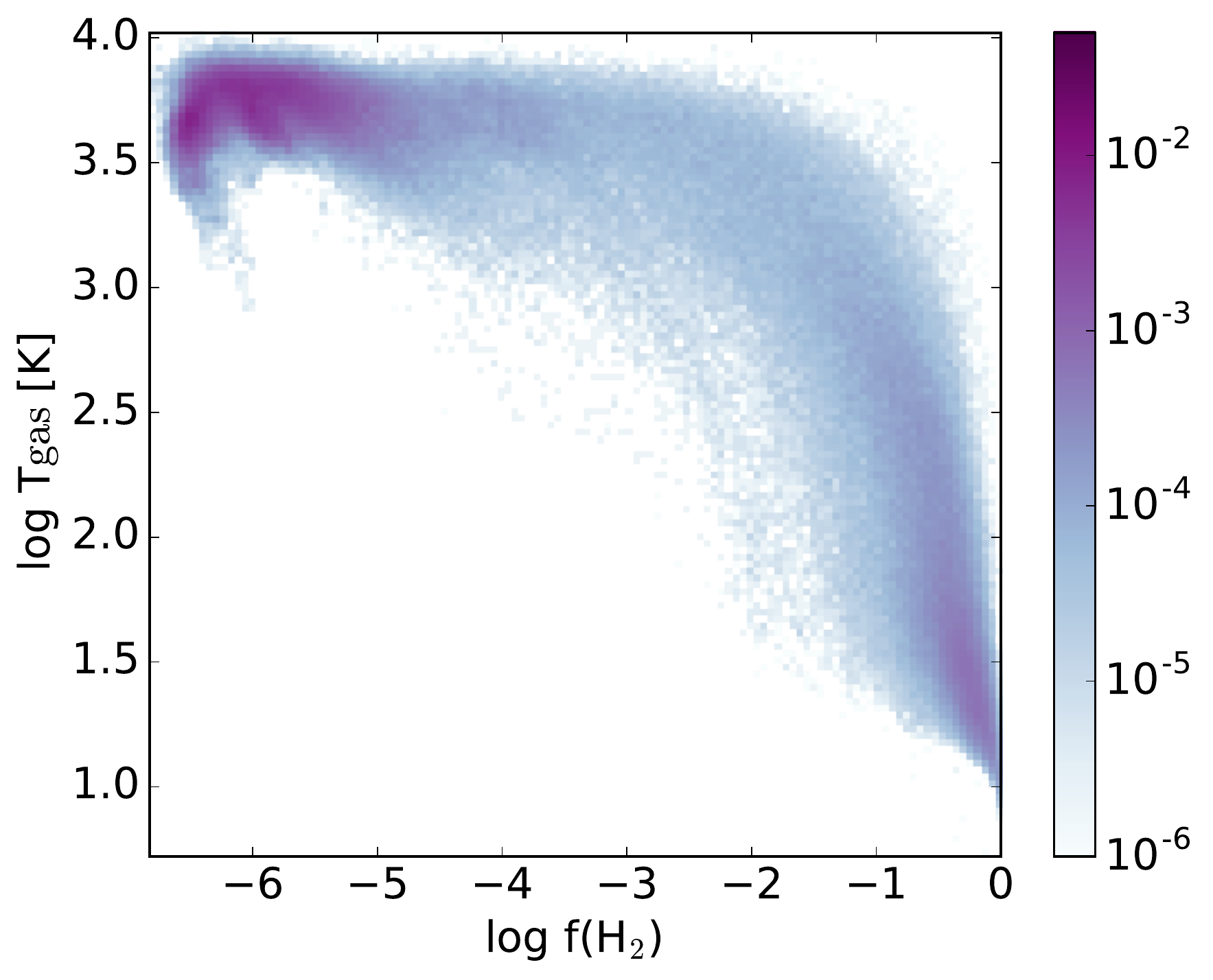}
\includegraphics[width=6cm]{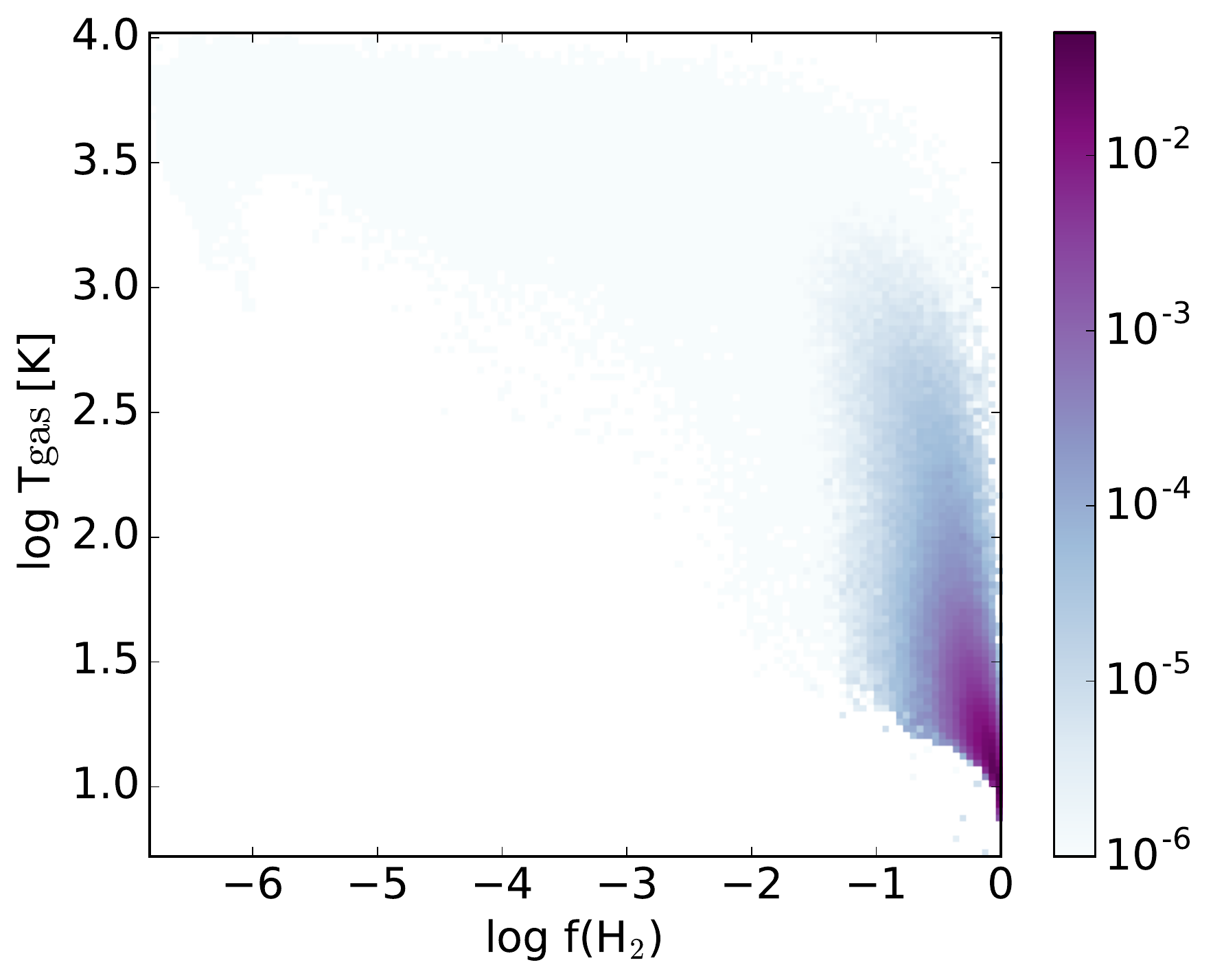}
\includegraphics[width=6cm]{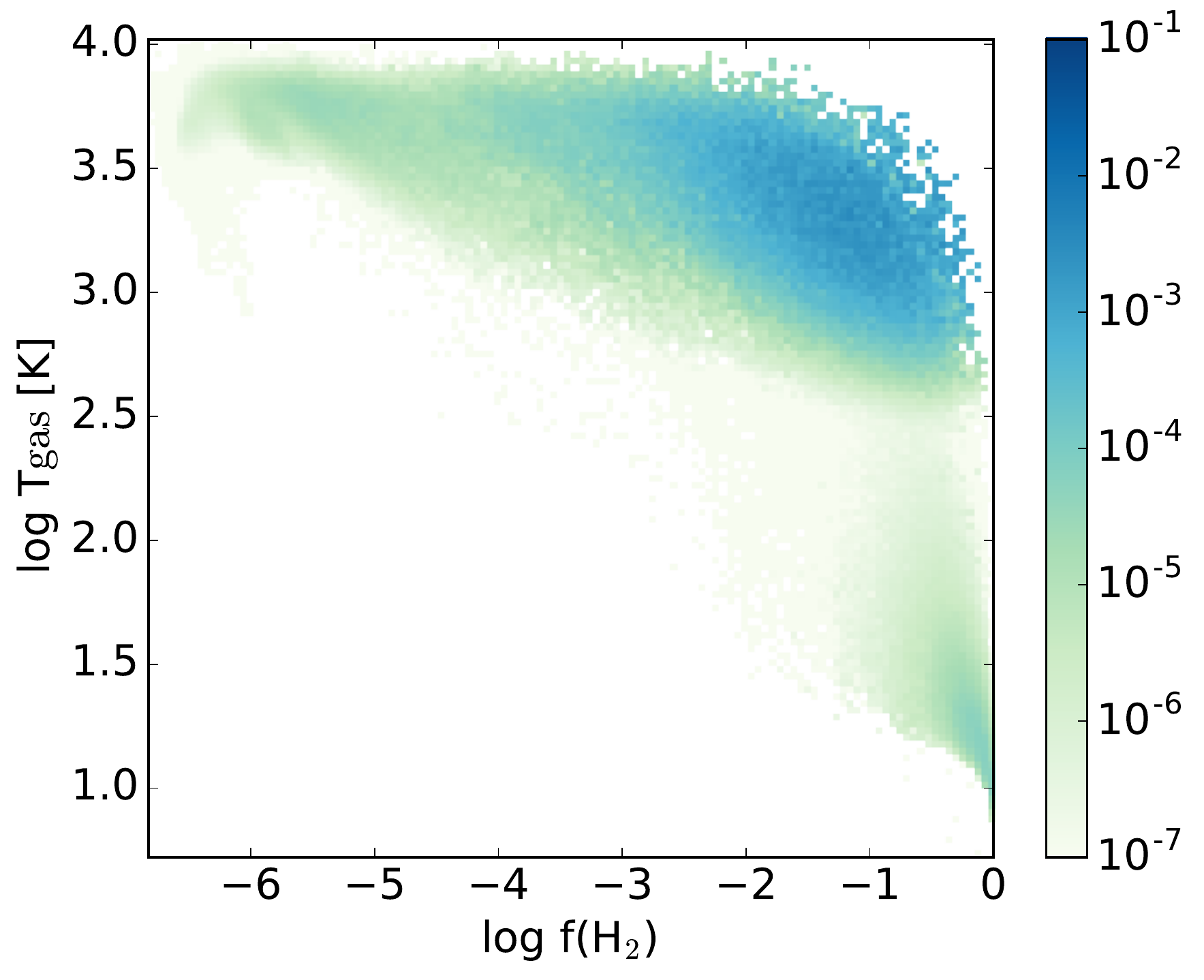}

\caption{ Volume-weighted (\emph{left panel}), $m(\mathrm{H}_2)$-weighted, and $m(\mathrm{CH^+})$-weighted 2D probability density functions (PDF) of the gas temperature ($T_\mathrm{gas}$) versus the H$_2$ fraction ($f(\mathrm{H_2})$).}
%\textbf{Normalised two-dimensional histograms representing respectively the gas volume} distribution (\emph{left panel)}, the H$_2$ \textbf{mass} distribution (\emph{central panel}), and CH$^+$ \textbf{mass} distribution (\emph{right panel}) as a function of the gas temperature and the H$_2$ fraction.} 
\label{H2dist_fig}
\end{figure*}
%%%%%%%%  FIG RESULTS %%%%%%%%%
%
%
%%
%=========================
\begin{center}
\begin{table}[h]
\caption{Momentum transfer coefficients.}
{\centering
\begin{tabular}{ l  l  l l }
\hline
\hline
Species $j$, $k$                        & $\mu_{jk}$ in $m_\mathrm{H}$          &       $K_{jk}$ ($\mathrm{cm^3 s^{-1}}$)                & Ref.\\
\hline
$\mathrm{C^+}, \ \mathrm{H}$            &       $12/13$         & $1.74\times10^{-9} v_\mathrm{rms}^{0.2}$           & ${(1)}$\\
$\mathrm{C^+}, \ \mathrm{H_2}$          &       $24/14$         & $0.79583\times K_{\mathrm{C^+}, \ \mathrm{H}}$         & ${(2)}$\\
$\mathrm{H^+}, \ \mathrm{H}$            &       $1/2$   & $2.13\times10^{-9} v_\mathrm{rms}^{0.75}$          & ${(1)}$\\
$\mathrm{H^+}, \ \mathrm{H_2}$          &       $2/3$   & $3.89 \times10^{-9} v_\mathrm{rms}^{-0.02}$         & ${(1)}$\\
$\mathrm{He^+}, \ \mathrm{H}$           &       $4/5$   & $2.58 \times10^{-9}$                                                  & ${(3)}$\\
$\mathrm{He^+}, \ \mathrm{H_2}$ &       $8/6$   & $2.17 \times10^{-9}$                                                  & ${(3)}$\\
$\mathrm{S^+}, \ \mathrm{H}$            &       $32/33$         & $2.34 \times10^{-9}$                                                  & ${(3)}$\\
$\mathrm{S^+}, \ \mathrm{H_2}$          &       $64/34$         & $1.98 \times10^{-9}$                                                  & ${(3)}$\\
\hline
\end{tabular}\\
\tablebib{
   (1)~\citet{pinto2008}; (2)~Assuming the same ratio as for the Langevin rates; (3)~Langevin rates (provided by A. Faure, priv. communication).
   }
%($a$) From \citet{pinto2008}\\
%($b$) Langevin rates (provided by A. Faure)
\par}
\label{tab:pintogalli}
\end{table}
\end{center}
%=========================
%
It is important to note  that $K_{jk}$, and thus the drift velocity $v_\mathrm{d}$, depends on the rms velocity $v_\mathrm{rms}$ (in $\mathrm{km\ s^{-1}}$). Assuming that all ions and neutrals have the same temperature, this velocity can be written as
\begin{equation}
v_\mathrm{rms} = \left(v_\mathrm{d}^2 + \frac{8 \ kT}{\pi \mu_{jk}}\right) ^{1/2} %\approx \left( \frac{8 \ kT}{\pi m_{jk}}\right) ^{1/2}
\label{eq:velrms}
,\end{equation}

\noindent which is dependent on the ion-neutral drift velocity itself, and on the thermal velocity \citep{pinto2008}. 

Since the effective temperature varies as the squared value of the drift velocity (see Eq.~\ref{deltaT}), errors can be greatly amplified. For this reason we use an iterative method to give an accurate value of $v_\mathrm{d}$. 

%it is important As most of drift velocities are weak in comparison to the thermal velocities, we use an iterative method, that consists in a two-steps approximation. The first step is to neglect the drift velocity in Eq.~\ref{eq:velrms}, and use this value in the momentum transfer rates of Eq.~\ref{eq:driftvel}. Since the velocity drift also requires the values of the number densities of the ions and neutrals, we solve for the chemistry using the gas temperature to obtain a reasonable approximation of the chemical abundances. These abundances lead to a first guess of the drift velocity, which can be reinjected in Eq.~\ref{eq:velrms} to obtain the rms velocity $v_\mathrm{rms}$. Secondly, the rms velocity can be used to calculate an accurate value of the drift velocity, which is finally used in the chemistry solver to obtain the final chemical abundances. 

For the momentum transfer rate coefficients we use the accurate expressions given by \citet{pinto2008}, and the Langevin rates when a more accurate rate was not available. For the particular case of collisions between C$^+$ and H$_2$, we assumed a fraction of the value used for the collisions between C$^+$ and H, which corresponds to the ratio between the corresponding Langevin rates. In Table~\ref{tab:pintogalli} we summarise the interactions included in our calculations\footnote{We do not include collisions with He, which  further reduce the drift velocity distribution only slightly.}. 

% . In a second step, we reinject the drift velocity obtained into the $v_\mathrm{rms}$ and repeat the procedure. . We show in Appendix \ref{ap_vADcorr} that this correction is in general mild, except in the regions of strong drift velocities. 
%Gas temperature distributions  and the effect of the ion-neutral drift are analysed in Sect. \ref{sect_in_drift}.

 %A further approximation comes from the fact that  

%Equivalently $n_n$ and $n_i$ are the number densities of neutrals and ions. Finally, $m_\mathrm{H}$ is the mass of hydrogen.  $==>$ CHECK.

%In a recent paper \citet{myers2015} have analysed the role of the ambipolar diffusion on ion-neutral reactions. In their estimate they fixed the electron fraction to $x_\mathrm{e} \approx 10^{-4}$ 

%--------------------------------------------------------------------
\section{Results}\label{results}

In order to compare our data with available observations, we selected a uniform grid of 1024 lines of sight for which we solved for the chemistry and integrated for column densities. Each line of sight has been analysed using a constant numerical resolution of $1024$ cells in the  line of sight, or equivalently $\sim 0.05~\mathrm{pc}$, which corresponds to an intermediate resolution level of $\ell = 10$ for the native AMR resolution of the simulation.

\subsection{Role of warm H$_2$}

\begin{figure}[t]
\centering
\includegraphics[width=9.3cm]{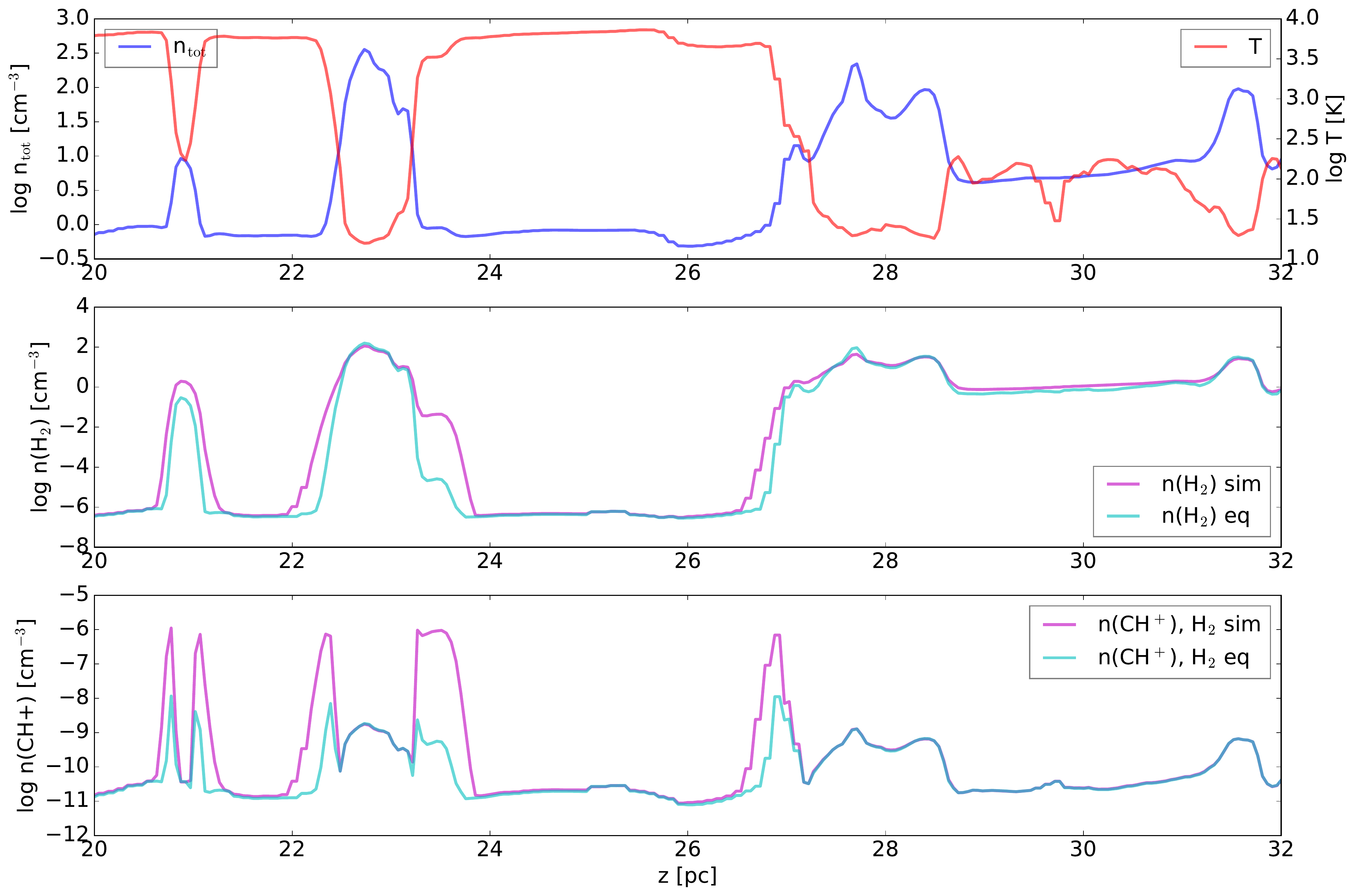}
\caption{Line of sight showing the influence of warm H$_2$. Each panel shows  the total number density and the gas temperature (\emph{top}), the number density of H$_2$ calculated at equilibrium and dynamically (\emph{centre}), and the number density of CH$^+$ (\emph{bottom}).} 
\label{LOS_NOFIX_fig}
\end{figure}

\begin{figure}[h]
\centering
\includegraphics[width=8.cm]{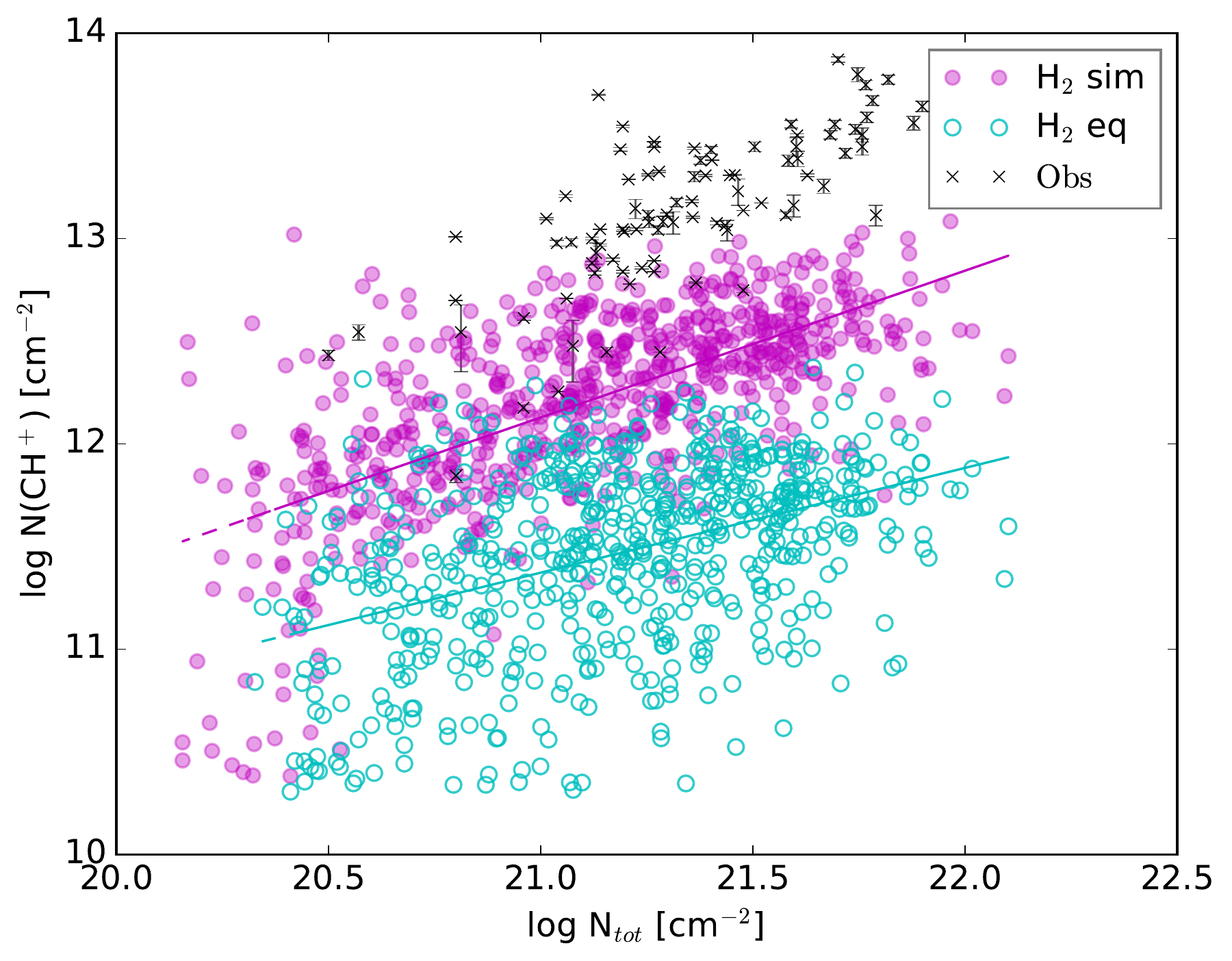}
\caption{Comparison of column densities of CH$^+$ as a function of the total column density for the case where H$_2$ is fixed from the simulation (\emph{solid circles}), and for the case where H$_2$ is calculated at equilibrium (\emph{open circles}). The crosses are the observational data from \citet{crane1995, gredel1997}, and \citet{weselak2008}.}
\label{NCHpNOFIX_fig}
\end{figure}

In \citetalias{valdivia2016} we  showed the presence of H$_2$, which is not at chemical equilibrium, in warm regions. As H$_2$ is a prerequisite to the formation of other molecules, and as warm reactants can boost the formation of molecules with high reaction barriers, we analyse the role of warm H$_2$ on the formation of CH$^+$. For this purpose we calculate the formation of CH$^+$ in two different ways.
In the first case we calculate the chemical abundances of all the species in our chemical network with the H$_2$ abundances obtained in our simulation (out of equilibrium H$_2$), while in the second case we calculate the full chemical equilibrium for all the species including H$_2$. %For each case we calculate the effective temperature consistently.

Figure \ref{H2dist_fig} shows the 2D probability density functions (PDF) in the form of a 2D histogram of the gas volume,  the H$_2$ mass ($m(\mathrm{H_2})$), and the CH$^+$ mass ($m(\mathrm{CH^+})$) distributions in the simulation box. The left  and centre panels show that a large number of cells are characterised by low densities of H$_2$, while a reduced number of cells concentrate most of the mass in H$_2$ form. The right panel shows that in spite of this distribution, most of the CH$^+$ is produced in regions with intermediate H$_2$ fractions ($f(\mathrm{H_2})\sim 0.3 - 30\%$), which do not correspond to the regions that dominate the volume or the mass of H$_2$, and with gas temperatures as high as several $10^2-10^3~\mathrm{K}$.     

We present a single line of sight in Fig.~\ref{LOS_NOFIX_fig} to shed some light on the physical conditions that give rise to an enhancement on the CH$^+$ abundance. This figure shows the local physical condition of the gas, as well as a comparison of the H$_2$ density calculated dynamically in our simulation and that expected at equilibrium for the same physical conditions (total density, temperature, dust shielding, and H$_2$ self-shielding). The bottom panel of this figure shows that most of the CH$^+$ is produced in regions that present specific characteristics  favourable to CH$^+$ formation: the fraction of H$_2$ is higher than is predicted at equilibrium, and temperatures are  of the order of several $100~\mathrm{K}$. In both cases most of the CH$^+$ is produced near the edge of the clumps, but the abundance of CH$^+$ is up to three orders of magnitude larger when the warm H$_2$ is used.   

The resulting column densities for the grid of 1024 lines of sight are shown in Fig. \ref{NCHpNOFIX_fig}. The column densities of CH$^+$ are $3 - 10$ times higher than those obtained with H$_2$ at equilibrium and closer to the observed ones. However, the abundances are still underpredicted by a factor of $\sim6$. This means that the warm H$_2$ plays an important role in the production of CH$^+$, although it is not enough to explain the observed abundances.

\subsection{Role of the ion-neutral drift} \label{sect_in_drift}

%%%%%%%%%%%%%%%%%%%%%%%%%%%%%%%%%%
%\footnote{P stands for probability distribution.} 
\begin{figure}
\centering
\includegraphics[width=7.8cm]{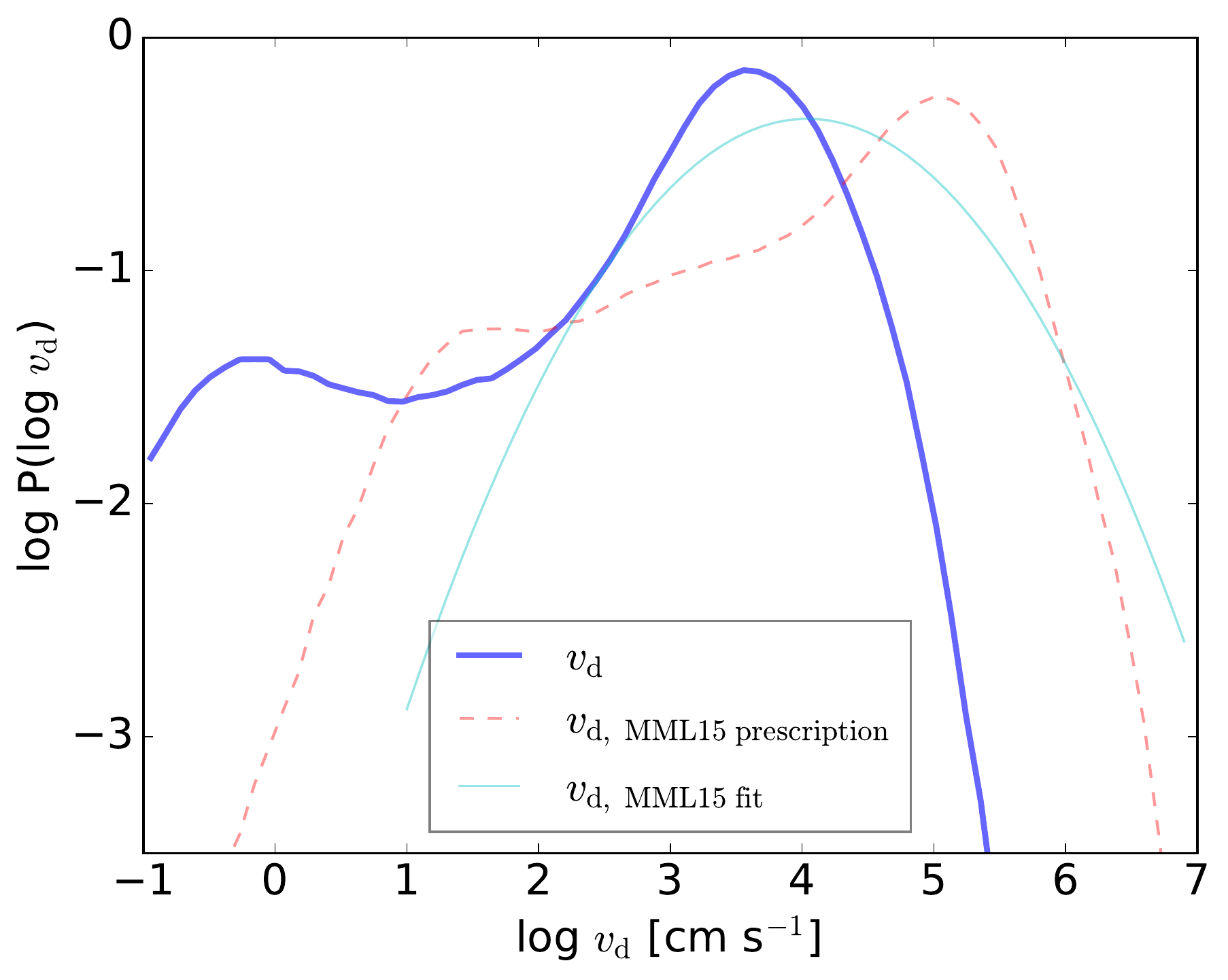}
%\includegraphics[width=7.8cm]{Figures/F8_9_11_15Myr_il10_los5_vAD_dist.pdf}
%\caption[Caption for LOF]{Real caption\footnotemark}
\caption[Caption for LOF]{Normalised volume-weighted drift velocity probability distribution ($P$), calculated as described in Sect.~\ref{indrift} (\emph{solid blue line}) compared to what would be obtained using the $\gamma_\mathrm{AD}$ and $x(\mathrm{e})$ prescriptions from \citetalias{myers2015} (\emph{dashed red line}). The analytic fit of \citetalias{myers2015} is given for comparison (\emph{solid light blue line}). }
\label{vADcorr_dist_fig}
\end{figure}

%%%%%%%%%%%%%%%%%%%%%%%%%%%%%%%%%%
\begin{figure}
\centering
\includegraphics[width=7.8cm]{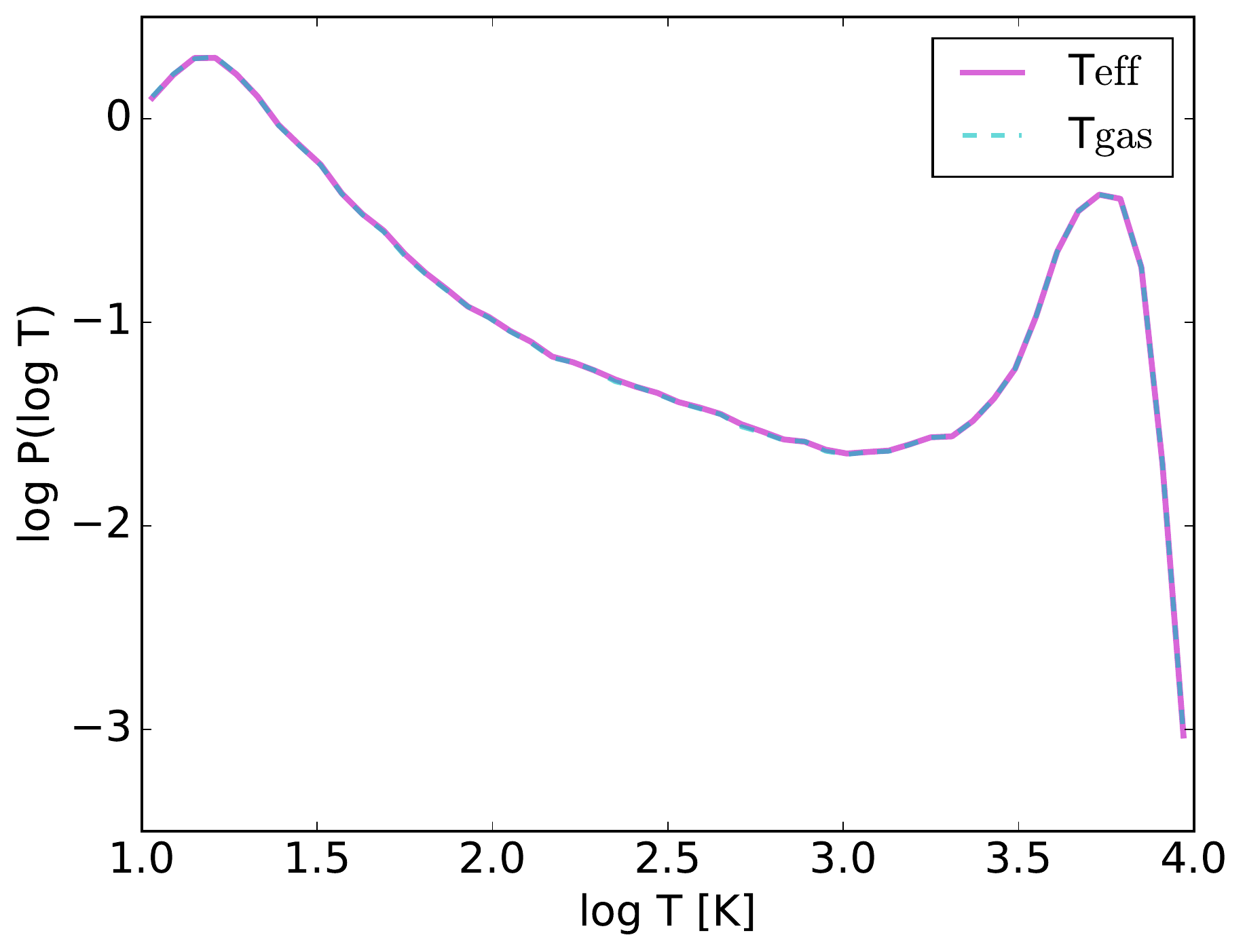}
\caption{Normalised mass-weighted probability distribution ($P$) of the gas temperature, $\log T_\mathrm{gas}$ (\emph{dashed light blue line}), and the effective temperature, $\log T_\mathrm{eff}$ (\emph{solid magenta line}), showing the two phases of the ISM.}
\label{Tdist}
\end{figure}

%%%%%%%%%%%%%%%%%%%%%%%%%%%%%%%%%%
%%%%%%%% FIG ION NEUTRAL DRIFT %%%%%%%%%
%\begin{figure*}[t]
%\centering
%\includegraphics[width=1.\textwidth]{Figures/STAT_Gcorr_9_11_15Myr_Teff_il_10los_4_stat_cells_0_histo2d_Phys_massW_all.pdf}\\
%\includegraphics[width=1.\textwidth]{Figures/STAT_Gcorr_9_11_15Myr_Teff_il_10los_4_stat_cells_0_histo2d_MML_all.pdf}
%\caption{Gas mass distribution (\emph{top panels}) in blue-purple shades, and CH$^+$  (\emph{bottom panels}) in green-blue shades. The color scale corresponds to the respective mass fraction. }
%\label{phys_stats_fig}
%\end{figure*}
%%%%%%%% FIG ION NEUTRAL DRIFT %%%%%%%%% 

The relative velocity between ions and neutrals adds a non-thermal component to the gas temperature that increases the effective temperature at which ion-neutral reactions occur. To assess the role of the ambipolar diffusion on the production of CH$^+$, we first analyse the distribution of ion-neutral drift velocities and then its impact on the effective temperature distributions.

\subsubsection{Drift velocity distributions}

Figure~\ref{vADcorr_dist_fig} shows the distribution obtained by \citetalias{myers2015} represented by their analytic fit (light blue line); the distribution of the drift velocity that we obtain using their prescription, which assumes that the dominant ion is C$^+$, with a constant abundance of $x(\mathrm{C^+}) = 1.6\times 10^{-4}$ and an ambipolar diffusion coefficient given by $\gamma = 8.47\times 10^{13}~\mathrm{cm^{3} s^{-1} g^{-1}}$ \citep{draine1980}, applied to our data (dashed red line); and the distribution obtained for our data using our prescription  as described in Sect.~\ref{indrift} (solid blue line).  

When we apply the prescription of \citetalias{myers2015} to our data (dashed red line), the distribution we obtain is different  to their distribution. We obtain an asymmetric distribution, narrower around the maximum, with a flatter region at low drift velocities and displaying a sharply decaying tail at higher drift velocities. The maximum, when using our simulation data, is located at $\sim 10^5~\mathrm{cm\ s^{-1}}$, which is one order of magnitude higher than the \citetalias{myers2015} distribution. Nevertheless, our distribution predicts  fewer  events (roughly one order of magnitude) of velocities $v_\mathrm{d} \gtrsim 10~\mathrm{km \ s^{-1}}$. The difference in shape might be a consequence of the biphasic nature of our simulation, while the higher peak can be easily explained by the different mean densities $\bar{n}$ of the simulated clouds. As in a first approximation the drift velocity is inversely proportional to the squared density (see Eq.~\ref{eq_vAD}); a mean density of $\bar{n}=30~\mathrm{cm^{-3}}$ in the case of \citetalias{myers2015} and $\bar{n}=9~\mathrm{cm^{-3}}$ in our case leads to a difference of roughly a factor of 10.

When we use our prescription (using the iterative method and the ion densities calculated with our chemical solver)  we obtain a distribution shifted towards lower drift velocities. The maximum of our distribution is located around $\sim 4\times10^3~\mathrm{cm\ s^{-1}}$, and most of the points in the simulation show very weak drift velocities. Points with drift velocities higher than $ 1~\mathrm{km \ s^{-1}}$ are extremely rare. These differences can be explained by the different assumptions we used. In the prescription of \citetalias{myers2015} the electronic fraction is assumed to be constant and dominated by the C$^+$ abundance, which is a valid assumption only in the dense and cold gas. This approximation underestimates the electronic fraction by one to two orders of magnitude in the more diffuse and warm gas where the electronic fraction is indeed dominated by the ionised hydrogen H$^+$ ($x_e \sim 10^{-2} - 10^{-3}$, as shown  in Fig.~\ref{LOS_vdrift_fig}). It is important to remember that $v_\mathrm{d} \propto 1/n_i$, and a difference of two orders of magnitude is amplified to four when calculating the effective temperature. In addition, the ambipolar coefficient $\gamma$ is assumed to be determined solely by interations between C$^+$ and H$_2$. 

%
%
% assumptions and by the different gas distribution. 

%%%%%%%%%%%%%%%%

%Our distribution is quite different from the one obtained by \citetalias{myers2015} (thin light blue line).  with a very asymmetrical distribution, being flatter towards smaller drift velocities, and sharply decaying towards high values.  

In summary, this figure shows that the simplified prescription of \citetalias{myers2015} applied to our data greatly overestimates the ion-neutral drift velocity distribution.

%%%%%%%%%%%%%%%%%

Figure~\ref{Tdist} shows the distribution of the gas temperature and the distribution of the effective temperature. It reveals the presence of the two distinct phases of the ISM, the warm neutral medium (WNM) and the cold neutral medium (CNM) \citep{field1969}, but also a transition phase that correspond to gas transiting through the thermally unstable phase between the two main phases. As the effect of the ambipolar diffusion is local and intermittent, it affects only a small fraction of the gas, and therefore the two phases are not expected to differ significantly in the effective temperature distribution. 
To noticeably change the relative distribution of the CNM and WNM, a mean drift velocity of at least $1~\mathrm{km\ s^{-1}}$ over the bulk of the CNM is necessary. 
%Furthermore, as the effect of the ambipolar diffusion is local and intermittent, these two phases are not expected to change. 
Differences should only appear in the intermediate temperature domain where the effective temperature is expected to vary  from the gas temperature for sufficiently high drift velocities ($v_\mathrm{d}\gtrsim 1~\mathrm{km\ s^{-1}}$). However, Fig.~\ref{Tdist} shows that the two distributions are almost indistinguishable, revealing that the effect of the ambipolar diffusion is negligible, at least under the conditions analysed in this work. %The effect is barely noticeable in the thermally unstable gas, which is located between the two main phases of the ISM. %Nevertheless, this contribution to the effective temperature can eventually help to overcome the reaction barrier in regions containing warm molecular hydrogen transiting through the thermally unstable phase.  
%==> IMPROVE 

\subsubsection{Role of ambipolar diffusion on the production of CH$^+$}

%As we recalled in the previous section, both the gas temperature and the ambipolar diffusion can increase the reaction rates (==> ???). To discriminate their relative importance  we present  in Fig.~\ref{phys_stats_fig} the total gas mass distribution and CH$^+$ mass distribution as two-dimensional histograms as functions of different parameters. The masses have been normalized to the total gas mass, and to the total CH$^+$ mass, respectively, for obtaining mass fractions.
% The first panels (top and bottom) of  Fig.~\ref{phys_stats_fig} show that most of drift velocities are comprised between $0.001~\mathrm{km\ s^{-1}}$ and $5~\mathrm{km\ s^{-1}}$. Nevertheless the production of CH$^+$ is mainly dependent on the gas temperature, and only at drift velocities higher than $\sim 3~\mathrm{km\ s^{-1}}$ the effect of the ambipolar diffusion becomes dominant. The second panels (top and bottom) show that CH$^+$ is mainly produced in low-density gas ($1 - 10~\mathrm{cm^{-3}}$), that is already warm ($300-5000~\mathrm{K}$). The third and fourth panels indicate that is for similar values of drift velocity, the production of CH$^+$ is mostly dependent on the H$_2$ fraction rather than on the drift velocity. 

%%%%%%%%%%%%%%%%%%%%%%%%%%%%%%%%%%
\begin{figure*}
\centering
\includegraphics[width=1.\textwidth]{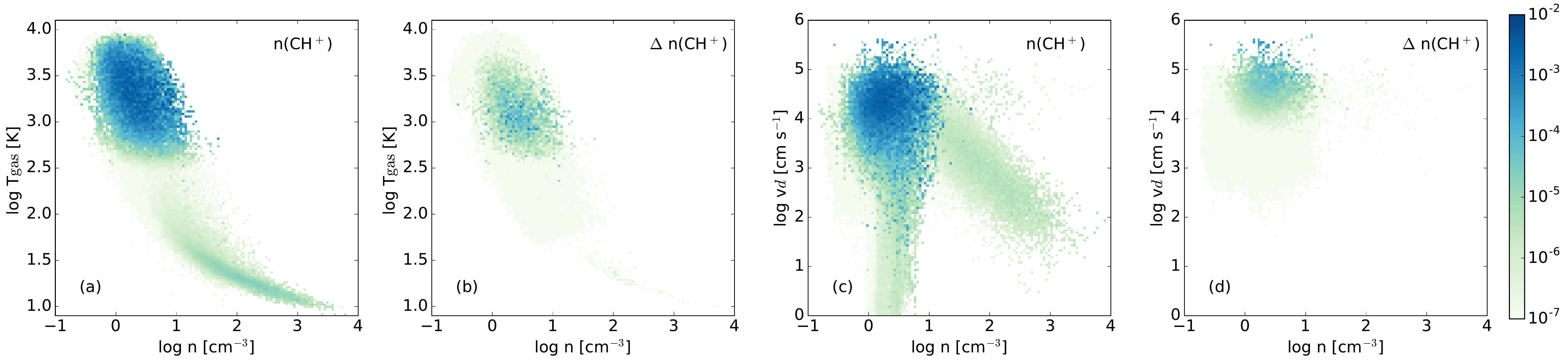}
\caption{Distribution of the abundance of CH$^+$ as a function of density and gas temperature (\emph{panel a}), and as a function of density and drift velocity (\emph{panel c}). \emph{Panels b} and \emph{d} show the contribution from the ambipolar diffusion.} \label{CHp_dist_fig}
\end{figure*}
%%%%%%%%%%%%%%%%%%%%%%%%%%%%%%%%%%

As we recalled in the previous section, both the gas temperature and the ion-neutral drift can increase the reaction rates.
To disentangle the role of the ion-neutral drift on the production of the CH$^+$ molecule, we calculated the equilibrium abundances of all the species (except for H and H$_2$) in two different ways. For the first case we calculated the chemical abundances using only the gas temperature, while in the second case we computed it taking into account the fact that the reaction rates depend on the ion-neutral drift velocity $v_\mathrm{d}$, as defined in Sect.~\ref{indrift}. With these two distributions we compute the excess produced by the ion-neutral drift as 
%
%the abundances of chemical species at equilibrium with the H$_2$ produced in the simulation 
%
%\begin{equation}
$\Delta n(\mathrm{CH^+}) = n(\mathrm{CH^+}) (T_\mathrm{eff}) - n(\mathrm{CH^+}) (T_\mathrm{gas}).$
%\end{equation}
%
%We first analyse the distribution of CH$^+$ for both cases. 

Figure~\ref{CHp_dist_fig} shows the distribution of CH$^+$ and the contribution from the ambipolar diffusion $\Delta n(\mathrm{CH^+})$ as 2D histograms. The two leftmost panels show the distributions as a function of the gas number density and gas temperature, while the two rightmost panels show the same distributions as a function of the gas number density and the ion-neutral drift velocity. This figure shows that the ion-neutral drift produces a small but not always negligible contribution to the total production of CH$^+$. A close comparison between the second and fourth panels shows that the production of CH$^+$  increases at temperatures ranging from several hundreds to thousands $\mathrm{K}$ and at low densities up to some $\sim 10~\mathrm{cm^{-3}}$ due to the ion-neutral drift, consistent with high values of the drift velocity at the same density regime.

\begin{figure}
\centering
\includegraphics[width=8.5cm]{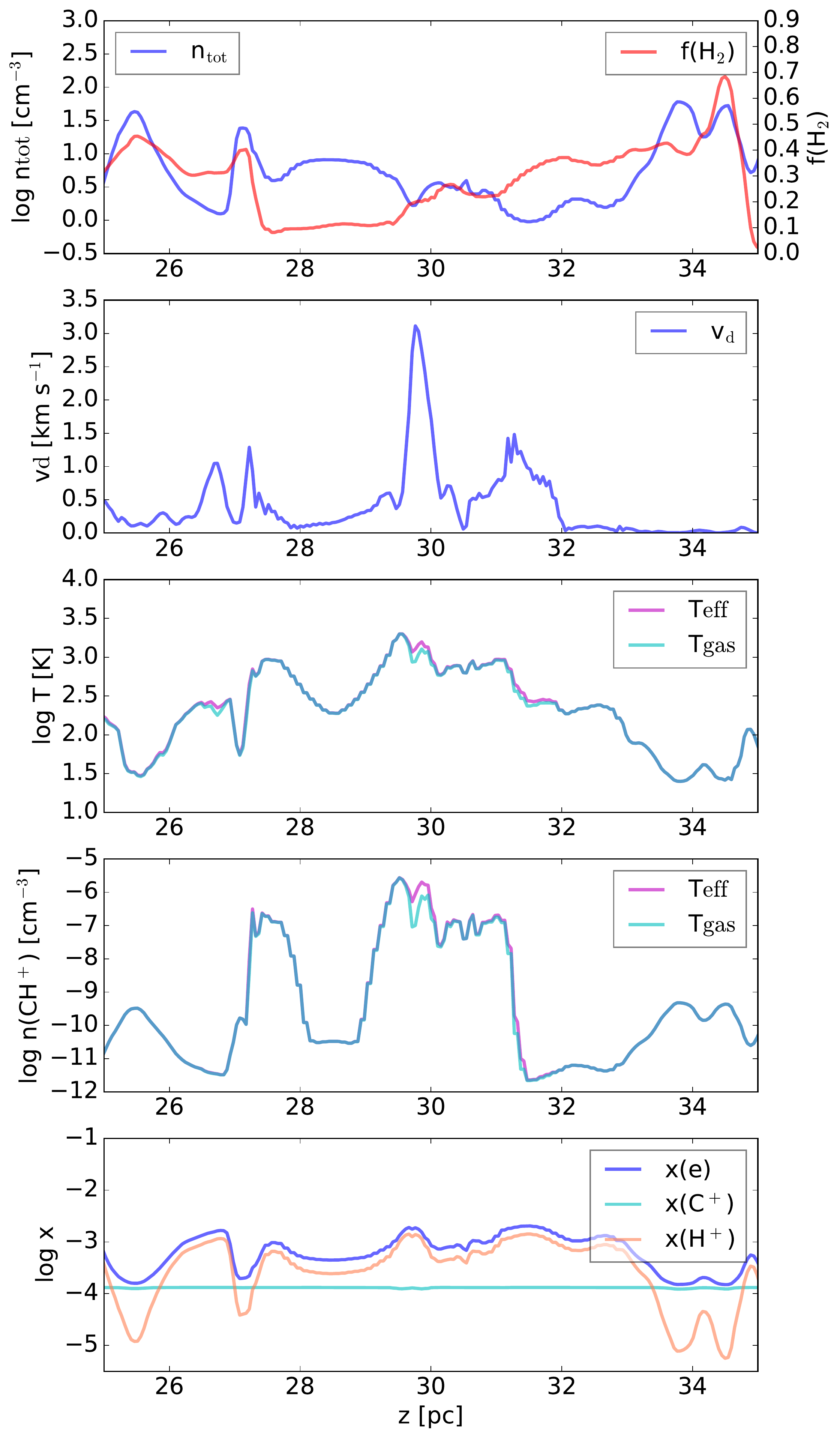}
\caption{Example of line of sight, showing a case of strong ion-neutral drift.}
\label{LOS_vdrift_fig}
\end{figure}

\begin{figure}
\centering
\includegraphics[width=8.cm]{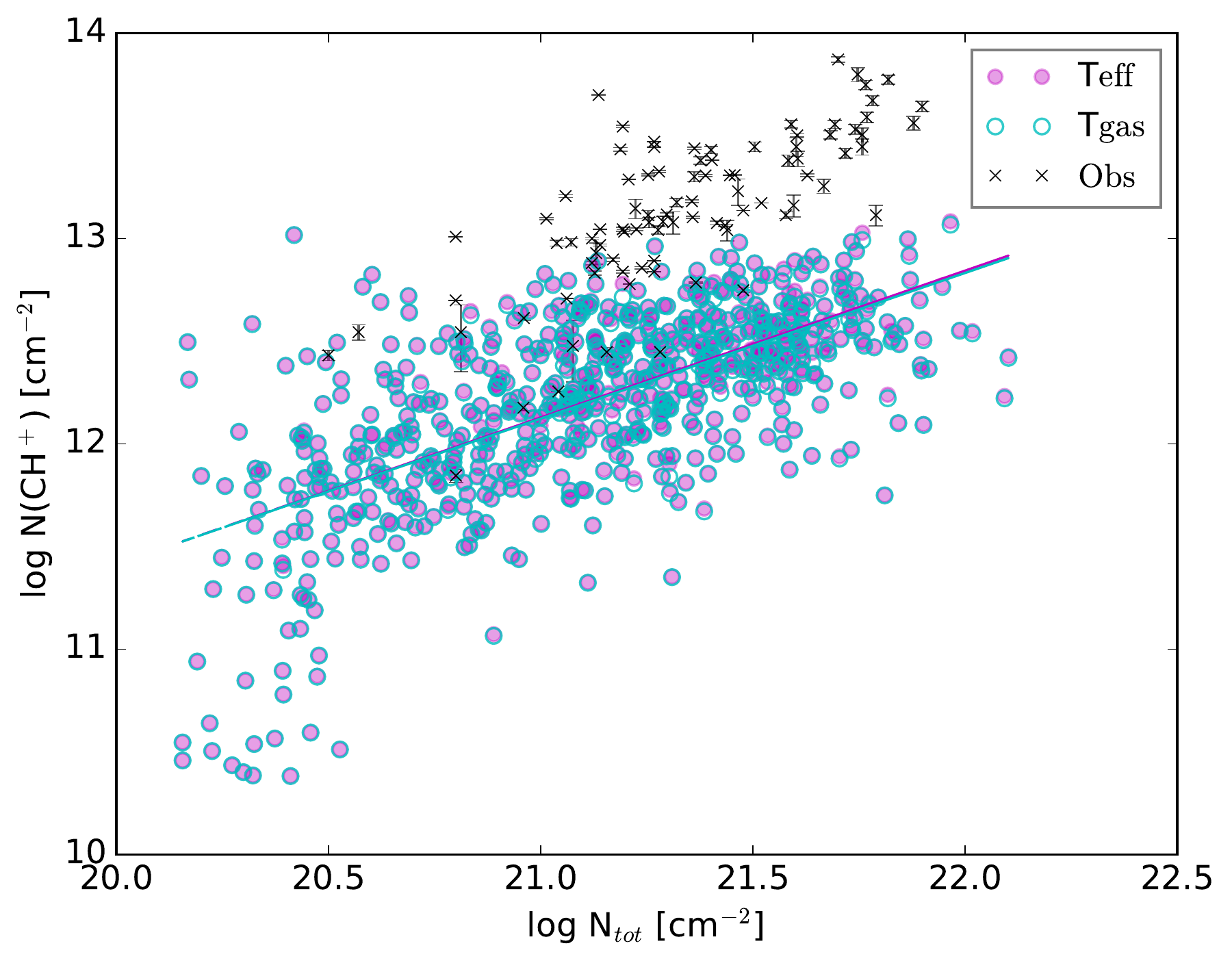}
\caption{Column densities of CH$^+$ as a function of the total column density. Including the ion-neutral drift (\emph{solid circles}), and without including this effect (\emph{open circles}). The crosses are the observational data as in Fig.~\ref{NCHpNOFIX_fig}.}
\label{NcolCHpTgasTeff_fig}
\end{figure}

As shown in Fig.~\ref{vADcorr_dist_fig} and in Fig.~\ref{Tdist} the effect of the ion-neutral drift is in general minor, but some regions of high drift velocities can arise. This is the case in Fig.~\ref{LOS_vdrift_fig}, where we selected a line of sight that displays a case of strong ion-neutral drift. This line of sight shows that high drift velocities arise as narrow features in regions with low density, where the electron abundance is in general higher than $10^{-3}$, and dominated by H$^+$ rather than by C$^+$. The production of CH$^+$ is locally strongly boosted (by $\sim 1$ order of magnitude), and although punctual and rare, such local events might have a great influence  on  the integrated column density. To check whether the ambipolar diffusion can have a significant impact on the global picture or if the effect is statistically unimportant, we integrate total column densities and CH$^+$ column densities along the selected lines of sight for the two cases analysed in this section. The global effect of the drift velocity is shown in Fig.~\ref{NcolCHpTgasTeff_fig}. This figure shows that, statistically, the effect of the ambipolar diffusion is almost imperceptible. Nevertheless, this effect is resolution-dependent, so probably the relevant scales for the ambipolar diffusion are underdescribed. This effect is explored in the following section.

%%%%%%%%%%%%%%%%%%%%%%%%%%%%%%%%%%
%%%%%%%%%%%%%%%%%%%%%%%%%%%%%%%%%%
%%%%%%%%%%%%%%%%%%%%%%%%%%%%%%%%%%

%\subsection{Physical conditions that give rise to an enhanced abundance of CH$^+$}

\subsection{Effect of the resolution on the $v_\mathrm{d}$ distribution}
\label{resolution}

%==> WE KNOW ALREADY THAT THE AD LENGTHSCALE IS SMALLER

Our estimate of $v_\mathrm{d}$ from the ideal MHD simulations depends
on, among other variables,    the gradient of the magnetic field and the
ion-neutral momentum transfer rate (sensitive to both the ionisation degree and 
the temperature). Both of these quantities are likely to be sensitive to
the resolution, so it is important to analyse the role of the
numerical resolution on the distribution of drift velocities.
%
% it is important to analyse the . In particular Fig.~\ref{NcolCHpTgasTeff_fig} shows that the correction induced by the drift velocity is important at high column densities, where densities are also high and thus numerical resolutions are high too (it is worth to remind that the refinement criteria is a density-based one). 

\begin{figure}
\centering
\includegraphics[width=7.8cm]{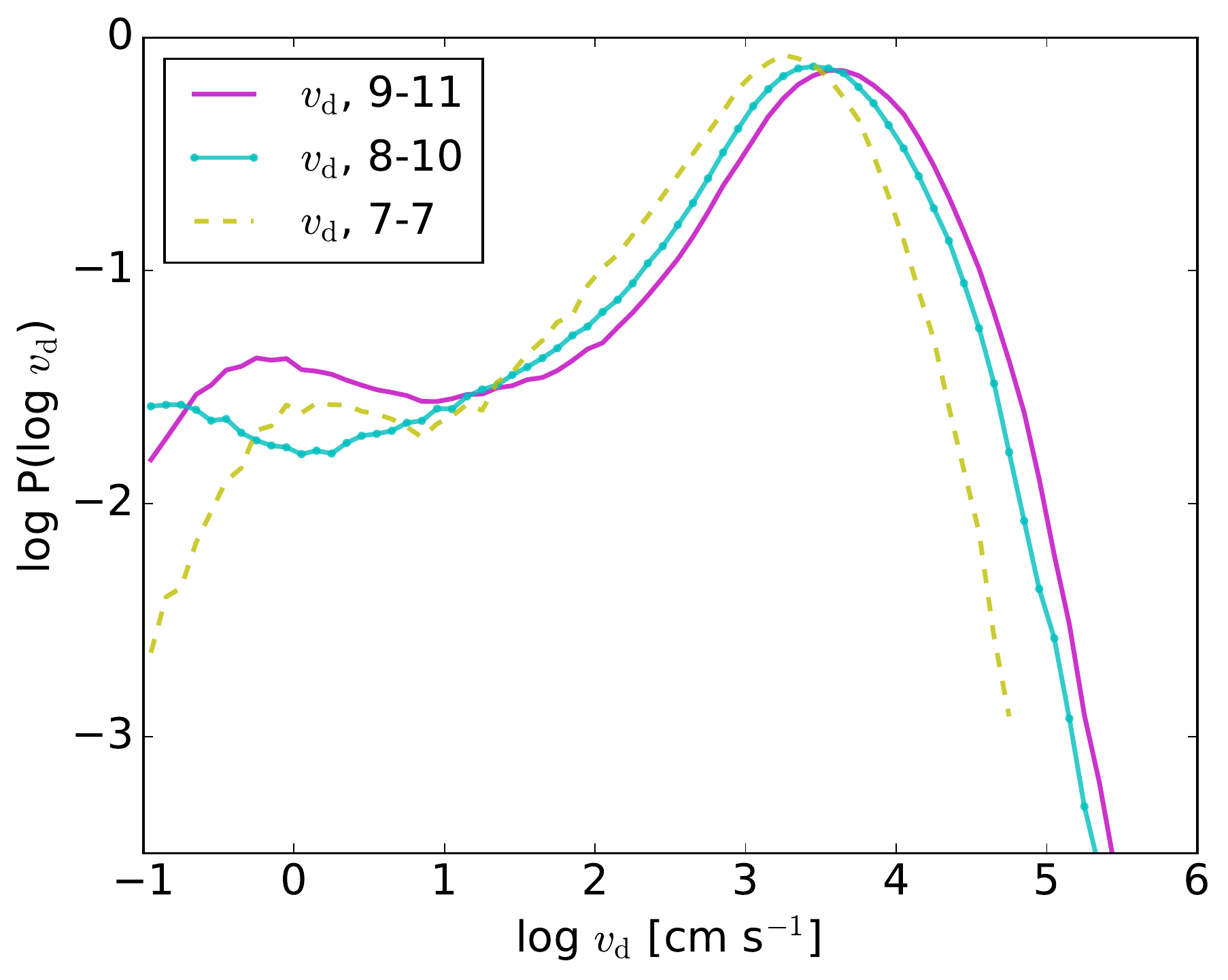}
\caption{Normalised probability distribution ($P$) of the ion-neutral drift velocity estimated from ideal MHD simulations at different numerical resolutions. Numbers in the legend indicate the lowest and highest resolution level ($\ell$) used for the AMR.}
\label{vAD_RES_fig}
\end{figure}

To shed some light on the role of  resolution, we present in Fig.~\ref{vAD_RES_fig} the distribution of drift velocities for three identical simulations, at $15~\mathrm{Myr}$, performed using three different numerical resolutions. Each simulation is characterised by the minimum ($\ell_\mathrm{min}$) and maximum ($\ell_\mathrm{max}$) resolution levels allowed in the AMR\footnote{The associated resolution for a level $\ell$ is $\mathrm{d}x = L/2^{\ell}$}, and by the constant intermediate resolution ($\mathrm{d}x_\mathrm{a}$) used for the analysis. The low-resolution simulation has $\ell_\mathrm{min} = \ell_\mathrm{max} = 7$, and $\mathrm{d}x_\mathrm{a} \simeq 0.4~\mathrm{pc}$; the intermediate-resolution simulation has $\ell_\mathrm{min} =8$, $\ell_\mathrm{max} = 10$, and $\mathrm{d}x_\mathrm{a} \simeq 0.1~\mathrm{pc}$; and the high-resolution simulation has $\ell_\mathrm{min} =9$, $\ell_\mathrm{max} = 11$, and $\mathrm{d}x_\mathrm{a} \simeq 0.05~\mathrm{pc}$.

%(to calculate the magnetic field gradients and drift velocities)

%To shed some light on the role of  resolution, we present in Fig.~\ref{vAD_RES_fig} the distribution of drift velocities for three identical simulations, at $15~\mathrm{Myr}$, performed using three different numerical resolutions. Each one is characterised by the minimal ($\ell_\mathrm{min}$) and maximal ($\ell_\mathrm{max}$) resolution levels allowed in the AMR, as well as the constant intermediate resolution ($\mathrm{d}x_\mathrm{a}$) used for the analysis, given in Table~\ref{tab:simres}. The low-resolution simulation has a uniform resolution of $\sim 0.4~\mathrm{pc}$, the intermediate one has an effective maximum resolution of $\sim0.05~\mathrm{pc}$, and the high-resolution simulation has an effective maximum resolution of $\sim0.025~\mathrm{pc}$. 
%%(to calculate the magnetic field gradients and drift velocities)
%%=========================
%\begin{center}
%\begin{table}[h]
%\caption{Numerical parameters of the simulations}
%{\centering
%\begin{tabular}{ l  l  l  l }
%\hline
%\hline
%Simulation     & $\ell_\mathrm{min}$   &       $\ell_\mathrm{max}$     &       $\mathrm{d}x_\mathrm{a}$ [pc]\\
%\hline
%Low                    &       $7$             &        $7$    &       $0.391$\\
%Intermediate   &       $8$             &        $10$   &       $0.098$\\
%High                   &       $9$             &        $11$   &       $0.049$\\
%\hline
%\end{tabular}
%\par}
%\label{tab:simres}
%\end{table}
%\end{center}
%%=========================

Figure~\ref{vAD_RES_fig} shows the distribution of the drift velocity obtained for each  of the simulations.
%the native resolution of $\sim 0.4~\mathrm{pc}$ for the low-resolution one, while for the two simulations at higher resolution we fixed the resolution to $\sim 0.05~\mathrm{pc}$. ==> Should I use different resolutions in the analysis? 
%This figure shows a drift toward low velocities in the maximum of the distribution for lower resolution simulations (. 
%The fact that the distribution of drift velocities has not yet converged seems to confirm the hypothesis that since low resolution simulations are worse at capturing small-scale features, resulting in smoother gradients, the description of the scales at which the ambipolar diffusion happens are poorly described, producing the shifted distribution.
%Since low resolution simulations are worse at capturing small-scale features, resulting in smoother gradients, the distribution is shifted towards smaller drift velocitites.
Since in our MHD simulations the only dissipative process is numerical truncation, the magnetic current ($\nabla\times{B}$) power will pile up at smaller scales for higher resolution, which  can increase our estimate of $v_\mathrm{d}$. On the other hand, the numerical dissipation heating will also occur at smaller scales as the resolution increases, which  will impact the temperature and the ionisation degree and may alter the momentum transfer rates in a way which is not easy to predict. The current resolution study indicates that the net effect is to shift the drift velocity distributions towards larger values for higher resolution simulations.

  If the ambipolar diffusion is included in the MHD simulations, this
  implies   accurately resolving the local ambipolar diffusion
  scale $\lambda_{\rm d}=\pi v_A/ (\gamma \rho_i)$, where $v_A$ is the
  Alfvén speed \citep{ntormousi2016, hennebelle2013}. This can be very
  stringent, although \citep{momferratos2014} has shown that the largest
  drift velocities can be obtained for scales much larger than
  $\lambda_{\rm d}$. In addition, observational constraints, based on
  kinematical observations of ionic and neutral species, set the drift
  scales at mpc-scales \citep{hezareh2014, li2008}.  Future studies
  should therefore strive to treat  the dissipative processes more accurately,
  but it is not clear how this will affect our results.

%==> SAY MORE

%The numerical resolution reveals itself as a key parameter on the description of the ambipolar diffusion. With increasing resolution, more cells with higher drift velocities appear, likely due to an improved description of small-scale gradients. The ion-neutral drift can therefore reach higher values and play a significant role in the reactions between neutrals and ions. 

%  ==> IMPROVE  : we need more resolution
%
%indicates that the resolution is a key parameter on the description ...

%{Distribution of the ion-neutral drift velocities and temperature}

%\begin{figure}
%\centering
%\includegraphics[width=8.cm]{Figures/NEW_highvAD_bis7_7_7_15Myr_Teff_H2fix_vd_il_7los_4_NcolsCOMP_RES_Teff_paper_CHp.pdf}
%\caption{Distribution of $\log v_\mathrm{d}~\mathrm{[cm\ s^{-1}]}$ for simulations at different numerical resolutions. Numbers in the legend indicate the lowest and highest resolution used for the AMR.}
%\label{Ncol_RES_fig}
%\end{figure}

%\subsection{Influence of the choice of $\gamma$ on the $v_\mathrm{d}$ }
% Line-of-sight analysis

%\subsection{Comparison to observations}

%\subsection{Comparison to other theoretical/numerical works }
%==> Myers2015
%-arguments
%$\gamma$
%$\zeta$
%$fH2$

\section{Conclusions}\label{conclusions}

%==> General conclusions...
We assessed the importance of two key factors that contribute to producing the methylidine cation CH$^+$ in diffuse molecular clouds, namely the role of warm reactants and the role of ambipolar diffusion. To this end, we  post-processed an ideal MHD simulation of a realistic turbulent molecular cloud, which includes the evolution of the thermal state of the gas and the formation of molecular hydrogen. For the post-processing we developed a chemical solver able to calculate the chemical equilibrium for a given set of physical conditions and able to fix the abundance of H$_2$ as an input parameter if required. The chemical solver provides the abundances of all the species in the chemical network (149 species in our case) very efficiently.

%==> On the role of warm H2
We compared the expected abundances of CH$^+$ obtained using a full equilibrium with those obtained when fixing the abundances of H$_2$ to the value obtained dynamically in the simulation. We  show that molecular hydrogen plays a fundamental role on the chemistry of the ISM. Since the evolution time of molecular hydrogen is long, its formation and evolution must be followed carefully in numerical simulations. In particular, we  show that the excess of H$_2$ found in warm gas as a consequence of the multiphase structure and the turbulent mixing within interstellar gas is a very important ingredient in the understanding of the warm chemistry.

More specifically, the formation of CH$^+$ seems to be more efficient in regions where H$_2$ is not expected at equilibrium. CH$^+$ is formed mainly in regions characterised by H$_2$ fractions close to $\sim 1-10~\%$ and by temperatures higher than $300~\mathrm{K}$, most likely of the order of $10^3~\mathrm{K}$. These specific regions, such as the external layers of clumps, meet the necessary conditions to allow endothermic reactions to produce CH$^+$ efficiently. Nevertheless, the abundances of CH$^+$ are still underpredicted compared to observations \citep{crane1995, gredel1997, weselak2008} by a factor of the order of $6$, suggesting that missing elements might contribute in the same measure. 
A possible clue to solving this puzzle, other than the dissipation of turbulence, is the excitation of rotational and vibrational levels of H$_2$ pumped by the surrounding UV field. Molecules of  H$_2$  transported towards regions where the shielding is mild are usually found in excited states, and the internal energy available in the molecule can be used to overcome the reaction barrier \citep{zanchet2013, herraez-aguilar2014}. It is also worth noting that given the timescales involved, the excited states are likely to be out of equilibrium and therefore need to be followed in real time in the simulation. 
A different but related issue would be the initial and boundary conditions used in our simulation. Since real molecular clouds are likely formed by multiphase flows made up of a mixture of WNM and CNM gas already containing molecular gas, the abundance of H$_2$ -- mainly in low-density regions -- might be underestimated. A larger fraction of H$_2$ in such regions might lead to an increase
in the CH$^+$ abundance.
%the formation of CH$^+$ would proceed at earlier stages in the evolution of the cloud. 
These effects are not treated in this paper, but we plan to address them in future works. 

%Among the missing effects we can name non thermal effects, other dissipation processes not described in our simulations, such as the ambipolar diffusion heating and the  turbulent heating.    

%the unexpected presence of molecular hydrogen in warm environments produces the necessary conditions to overcome endothermic reactions to produce CH$^+$ efficiently, as in the case of the border of clumps. 

%==> On the role of ambipolar diffusion
We calculated the abundances of the dominant ionic species using our full chemical network. This, along with accurate momentum transfer rates, leads to a distribution of drift velocities dominated by low values (peak around $\sim 4\times 10^3~\mathrm{cm\ s^{-1}}$), where drift velocities higher than $1~\mathrm{km\ s^{-1}}$ are extremely rare events. Our calculation of ion-neutral drift velocities does not produce a high-velocity tail, as has been shown in previous works \citep{myers2015}. This  highlights the importance of a good description of small-scale physical processes, such as the calculation of the electron density, as well as the description of the main ion-neutral interactions in order to avoid unrealistic estimates of the ion-neutral drift velocities. It is worth noting that since the corrective term $\Delta T$ depends quadratically on the drift velocity, and at the same time the formation rates are proportional to exponentials involving $\Delta T$, overestimated drift velocities can lead to an artificial overabundance of CH$^+$.   

%rare events. We show that a detailed calculation of the electronic fraction, the abundance of H2 and the momentum transfer rate is mandatory to obtain a reliable estimate of the ion-neutal velocity drift. 

%The effect of the expected distribution of drift velocities, in our case, is not able to produce a noticeable difference in the distribution of the effective temperature, and thus is not enough to explain the observed abundances of CH$^+$. Even though it can increase locally the abundance of CH$^+$, the role of the ambipolar diffusion is very limited, because high drift velocities are rare events. However, the comparison of results at different resolutions indicates that the scales of ion-neutral drift are not resolved yet and that smaller scales could have a strong influence on the high velocity tail. Therefore the ambipolar diffusion physics and its scales need to be taken into account in future works in order to describe this process and its impact on the production of interstellar CH+.

Our distribution of $v_\mathrm{d}$ does not  impact the effective temperature distribution, and thus is not able to  explain the observed abundances of CH$^+$. Even though it can increase locally the abundance of CH$^+$, the role of the ambipolar diffusion is very limited because high drift velocities are rare events. However, we show that resolution effects could have a strong influence on the high-velocity tail. Therefore, the ambipolar diffusion physics and its scales need to be taken into account in future works in order to describe this process and its impact on the production of interstellar CH$^+$.

%Is is worth mentioning that the resolution seems to be limiting the description of such small scale phenomenon, and future works must address this question.  

%==> Missing elements
%It is also worth mentioning that our simulations do not treat explicitly the dissipation of the mechanical energy, which occurs at very small-scales (at $\sim 40~\mathrm{AU}$ scales), and thus an additional and very localised heating could be at play. In the case of our simulation, if all the mechanical energy that is injected is disipated within the cloud, this will lead to an average turbulent rate of some $\dot\varepsilon \sim 10^{-25}~\mathrm{erg\ cm^{-3} s^{-1}}$, that could eventually trigger outbursts of CH$^+$ formation statistically significant.

%==> What else ?

\begin{acknowledgements}
We are grateful to Alexandre Faure, who kindly provided us with the Langevin rates, and to Edith Falgarone and Eva Ntormousi for valuable discussions.

We acknowledge the financial support of the Agence Nationale pour la Recherche through the COSMIS project. This research has received funding from the European Research Council under the European Community Seventh Framework Programme (FP7/2007-2013 Grant Agreement No. 306483).

We thank the French Programme Physique Chimie du Milieu Interstellaire (PCMI).

This work was granted access to HPC resources of CINES under the allocation x2014047023 made by GENCI (Grand Equipement National de Calcul Intensif). This work was granted access to the HPC resources of MesoPSL financed by the Region Ile de France and the project Equip@Meso (reference ANR-10-EQPX-29-01) of the programme Investissements d'Avenir supervised by the Agence Nationale pour la Recherche.

\end{acknowledgements}


\bibliographystyle{bibtex/aa} % style aa.bst 
\bibliography{biblio_val} % your references Yourfile.bib

\appendix

%=================================
\section{The chemical solver}
\label{ap_chemsolv}
The chemical solver built in this work is an adaptation of the solver used in the Meudon PDR 
(photodissociation regions) code \citep{lepetit2006}, modified to simplify the treatment of 
a few chemical processes (e.g. surface reactions, \HH\ self shielding), and optimised to 
improve the computational speed. The solver, which we describe in greater detail below, is 
available on the ISM numerical plateform of the Paris Observatory\footnote{\url{http://ism.obspm.fr}}.

\subsection{Description}

\begin{table}[!ht]
\begin{center}
\caption{Parameters of the chemical solver. The range of values found in the simulation
are given in Col. 3.}
\begin{tabular}{l l c l}
\hline
\multicolumn{4}{c}{mandatory} \\
\hline
$\chi$               & Mathis   & $1$              & external UV radiation field \\
$A_V$                & mag      & $0 - 10$          & visible extinction \\
$T_K$                & K        & $10 - 10^4$      & kinetic temperature \\
$\dens$              & \cc\     & $10^{-1} - 10^4$ & gas density \\
$\zetaHH$            & s$^{-1}$ & $3\times10^{-16}$       & CR ionisation rate of \HH\\
\hline
\multicolumn{4}{c}{optional} \\
\hline
$f_{\rm sh,\,\,\HH}$ &          & $10^{-8} - 1$  & \HH\ self-shielding factor$^a$ \\
$f_{\rm sh,\,\,CO}$  &          &               1 & CO   self-shielding factor$^b$ \\
$x(\HH)$             &          & $10^{-7} - 1$  & \HH\ abundance \\
$v_\mathrm{d}$              & \kms\    & $0 - 5$ & ion-neutral velocity drift\\
\hline
\end{tabular}
\begin{list}{}{}
($a$) \citet{valdivia2016}
($b$) not computed in the simulation
\end{list}
\label{Tab-Par}
\end{center}
\end{table}
The solver considers a static fluid cell of density \dens\ and temperature $T_K$, irradiated 
by an external radiation field $\chi$ (expressed in Mathis' unit, \citealt{mathis1983}) attenuated by a visible 
extinction $A_V$, and pervaded by cosmic ray particles. The cell contains $N_X$ species 
(resulting from the combinations of $N_A$ different atoms) that interact with each other 
through a given network of chemical reactions. In this configuration the code computes the 
chemical state of the fluid cell at equilibrium. Assuming that chemical reactions obey the 
law of mass action kinetics, this equilibrium is defined by the following set of algebraic 
equations
\label{EqChemSys}
\begin{equation}
\left.\begin{aligned}
& \frac{dn({\rm X}_i)}{dt} = \sum\limits_j^{N_{\mathcal{R}}} \left( \prod\limits_k n({\rm R}_{j,k}) \right) \,\,
  \kappa_j \,\, s_j({\rm X}_i) = 0 & \forall i \in [1:N_X] \\
& \sum\limits_j^{N_X} n({\rm X}_j) \,\, m({\rm A}_i,{\rm X}_j) = n_{{\rm A}_i} & \forall i \in [1:N_A] \\
& \sum\limits_j^{N_X} n({\rm X}_j) \,\, c({\rm X}_j) = 0 \\
\end{aligned}
\right.
\end{equation}
where $n({\rm X})$ and $n_{\rm A}$ are the density of species ${\rm X}$ and the number 
of atoms ${\rm A}$ (in \cc), $N_{\mathcal{R}}$ the number of reactions, ${\rm R}_{j,k}$ 
and $\kappa_j$ the reactants and reaction rate of reaction $j$, $s_j({\rm X})$ the 
stoichiometric coefficient of {\rm X} in reaction $j$, $m({\rm A},{\rm X})$ the 
multiplicity of atom ${\rm A}$ in species ${\rm X}$, and $c({\rm X})$ the charge of 
${\rm X}$. The chemical networks used in the solver are conservative; therefore,  the first line
above  provides a system of $N_X-N_A-1$ independent equations 
which is completed by conservation equations for each atom and for the electrons, 
chosen to replace the evolution equation of the most abundant atom carrier and charge 
carrier.

As described in \citet{lepetit2006}, the system of equations is solved with a Newton-Raphson 
scheme modified to prevent exploration of solutions with negative abundances. Iterations 
stop when the relative variation of abundances falls below a given threshold and if this 
solution corresponds to an equilibrium for the chemistry, i.e. when
\begin{equation}
\left.\begin{aligned}
& \sum\limits_i^{N_X} \frac{|\delta n({\rm X}_i)|}{n({\rm X}_i)} < \varepsilon_1 \\
& \sum\limits_i^{N_X} \frac{|F({\rm X}_i)-D({\rm X}_i)|}{F({\rm X}_i)+D({\rm X}_i)} < \varepsilon_2 \\
\end{aligned}
\right.
,\end{equation}
\label{EqChemCrit}
\noindent where $F({\rm X})$ and $D({\rm X})$ are the total formation and destruction rates of 
${\rm X}$ and $\varepsilon_1$ and $\varepsilon_2$ are two parameters set to $10^{-6}$ 
and $10^{-3}$, respectively. When called for the first time, the solver starts with 
initial conditions representative of the high-ionisation phase \citep{LeBourlot1995}: 
all carbon and sulfur in \Cp\ and \Sp, and all oxygen and nitrogen in O and N. When called 
in sequence, the initial conditions are the set of abundances corresponding to the last
solution found by the solver. This method is done to favour continuity of chemical states 
computed on adjacent fluid cells. However, it does not preclude the existence of one or 
several other chemically stable solutions for a given set of physical conditions (e.g. 
low-ionisation phase, \citealt{LeBourlot1995}).

The main input parameters of the solver and the ranges of values they span in the simulation
are given in Table \ref{Tab-Par}. In addition, the code accepts four optional input quantities:
the ion-neutral velocity drift (see Sect. \ref{indrift}), the self-shielding factors
for the photodissociation of \HH\ and CO (see eq. 11 of \citetalias{valdivia2016}), and the
fractional abundance of \HH\ (see Sect.~\ref{Sect-fixH2}).

\subsection{Chemical network}

The elemental abundances adopted in this work are set to the values observed in the solar
neighbourhood and compiled by \citet{Flower2003}, assuming no mantles on grain surfaces.
The chemical network is the most recent version of the network used in the Meudon PDR code, 
slightly adapted to simplify the treatment of a few chemical processes. This network is
available online (\url{http://ism.obspm.fr}, {\it network\_valdivia\_2016.dat}) and contains 
149 species interacting with each other through 2692 reactions.

Formation of \HH\ on grain surfaces and destruction by photodissociation are computed 
according to \citetalias{valdivia2016}. Contrary to the PDR model, which performs a 
coherent calculation of grain surface chemistry including both the Eley-Rideal and the 
Langmuir-Hinshelwood mechanisms \citep{lebourlot2012,bron2014}, the formation rate of 
\HH\ is computed via a simple function, scaled to the mean value of the formation rate 
observed in the ISM \citep{gry2002}, and tuned to depict the dependence of the sticking 
coefficient of H on grains on the kinetic temperature. The photodissociation rate of \HH\ 
is set to $3.3 \times 10^{-11} \chi f_{\rm sh,\,\,\HH}$ s$^{-1}$, where $f_{\rm sh,\,\HH}$ 
is a parameter (see Table \ref{Tab-Par}) used to include the effects of shielding of the 
radiation field by \HH\ line and dust continuous absorptions (e.g. 
\citealt{Draine1996}). 
%Similarly, the photodissociation rate of CO is set to $1.0 
%\times 10^{-10} \chi f_{\rm sh,\,CO}$ s$^{-1}$, where $f_{\rm sh,\,CO}$ is a parameter 
%used to take into account shielding by \HH\ lines and by dust particles 
%\citep{lee1996}. 
All other gas phase photoreaction rates are computed using the exponential 
fits of \citet{van-dishoeck1998} and \citet{van-dishoeck2006} at a given value of the 
visible extinction.

The treatment of the charges of polycyclic aromatic hydrocarbons (PAHs) and very small 
grains and of the processes of electron transfer between those particles and the ions 
is similar to the treatment performed in the Turbulent Dissipation Region (TDR) model 
\citep{godard2014}. PAHs and grains are described as spherical particles with radius 
of 6.4 \AA\ and 0.01 $\mu$m, respectively, and fractional abundances\footnote{Fractional abundances of PAHs and
grains  are computed assuming a dust-to-gas mass ratio of 0.01, 
a PAH mass fraction of 4.6 \% \citep{Draine2007}, an MRN size distribution of grains 
ranging from 0.005 to 0.3 $\mu$m, and a log-normal distribution of PAHs centred on
6.4 \AA.} of $4.2 \times 10^{-7}$ and $4.8 \times 10^{-10}$. 
Each particle is supposed to exist in charge states ranging from $-1$ to $5$. Photoionisations and photodetachments of all the corresponding species are modelled as photoreactions with rates deduced from detailed computations of the photoelectric effect with the Meudon PDR code. Electronic attachments and ion recombinations on dust (in their various charged states) are treated using the prescription of \citet{Draine1987} and \citet{Weingartner2001b,
Weingartner2001a}.

%Photoionisation of dust 
%particles are modelled as photoreactions in the gas phase with rates deduced from detailed 
%computations of the Meudon PDR code. Electronic attachment and ion recombinations on charged 
%dust are treated using the prescriptions of \citet{Draine1987} and \citet{Weingartner2001b,
%Weingartner2001a}.

\subsection{Fixed \HH\ abundance} \label{Sect-fixH2}

The 3D MHD simulation of the diffuse medium used in this work was built to compute 
the time dependent evolution of the abundance of molecular hydrogen. In order to estimate 
the impact of out-of-equilibrium \HH\ on the chemistry, the solver offers the possibility
of fixing the abundance of \HH\ in the cell. In this case, the code searches for a chemical 
state of the gas where almost all the species are at equilibrium.

This option is neither straightforward nor inconsequential. Fixing the abundance of 
\HH\ is equivalent to removing an equation from the first line of system \ref{EqChemSys}. 
System \ref{EqChemSys} therefore becomes an overdetermined problem composed of $N_X-1$ 
variables linked by $N_X$ independent equations. Consequently, if the abundance of \HH\ 
is fixed and out of equilibrium, either the system is not conservative or there is at 
least one other species out of equilibrium.

To simplify the problem, the solver works on the assumption that only one species,
${\rm S}_{\rm H}$, is out of equilibrium in addition to \HH. While this hypothesis is a good
approximation in some cases, it raises several issues: What
criteria should be used to select ${\rm S}_{\rm H}$? Does this choice guarantee the existence 
of a solution over a wide range of physical conditions (temperature, density, and molecular 
fraction)? If a solution exists, is it a correct representation of the ISM chemistry 
induced by out of equilibrium \HH?

\subsubsection{Selection criterion}

There are several arguments that constrain the selection of the out of equilibrium
species ${\rm S}_{\rm H}$. First, a chemical network used with a constant abundance of 
molecular hydrogen is conservative for all elements except H. A necessary condition 
to invert system \ref{EqChemSys} is therefore to remove an evolution equation of an 
H-bearing species. Secondly, the chemical equilibrium state of the ISM is driven by 
its most abundant constituents. In order to find a stable solution, ${\rm S}_{\rm H}$ 
should thus be abundant enough so that its density can be slightly adjusted to conserve 
the number of H atoms without modifying the at-equilibrium abundances of all the other 
species. Finally, since the timescale allowed to reach an equilibrium is limited 
by the resolution of the simulation, ${\rm S}_{\rm H}$ should be among the species 
which react the most quickly to any variation in the abundance of \HH, i.e. those most likely to be out of equilibrium if \HH\ is.

We therefore define ${\rm S}_{\rm H}$ as the most abundant H-bearing species besides \HH. 
With the parameters explored in the Appendix and the physical 
conditions spanned in the simulation (see Table \ref{Tab-Par}), we find that such a criterion 
always leads to atomic 
hydrogen (H), both in the most diffuse and ionised phases of the ISM ($\dens \sim 0.1$ \cc, 
$A_V < 10^{-6}$) and in the fully molecular gas. For intermediate or large molecular 
fractions ($f(\HH) \gtrsim 10^{-2}$), the selection method fulfils all the constraints. 
In particular, H is not only out of equilibrium, but 1D time-dependent chemical models 
also show that its abundance varies oppositely to that of \HH: $dn({\rm H})/dt = - dn(\HH)/dt$. 
In contrast, the method is found to be frailer at low molecular fraction ($f(\HH) \lesssim 
10^{-2}$). Indeed, in this case, the abundance of H is less dependent on that of \HH, 
hence several species are more likely than H to be out of equilibrium. Despite this 
limitation, we find that selecting H is paramount even in these conditions in order to 
ensure the numerical stability of the solver.

It is worth noting  that  H is the main species out of equilibrium in a large majority 
of cases, which strongly supports the method used in \citetalias{valdivia2016}: 
reducing the chemical network to H and \HH\ to compute the abundance of molecular
hydrogen.

\subsubsection{Existence of a solution}

\begin{figure*}[!ht]
\begin{center}
\includegraphics[width=1.\textwidth]{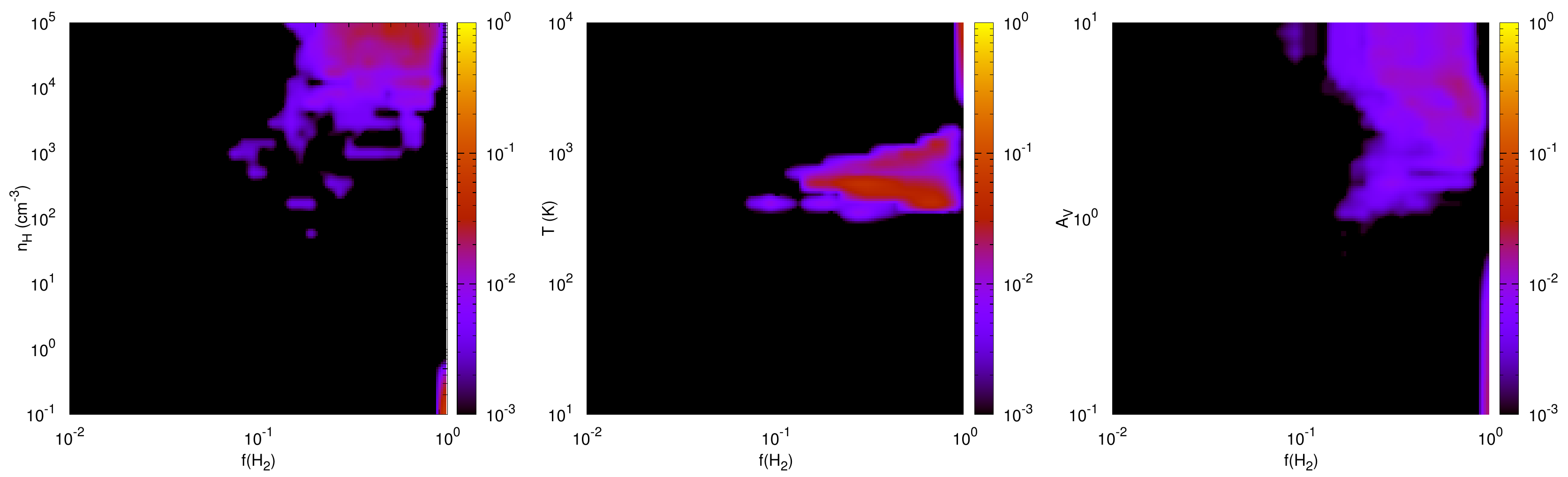}
\caption{Percentage of convergence failures of the chemical solver as functions of the 
molecular fraction and the density (\emph{left}), the temperature (\emph{middle}), or the visible 
extinction (\emph{right}).}
\label{Fig-fail}
\end{center}
\end{figure*}

Dynamical systems such as chemical networks under mass action kinetics are known to
display a variety of asymptotic behaviours depending on their control parameters (e.g. 
\citealt{vidal1988,feinberg1987}). In particular, fixing the abundance of one species, 
which is equivalent to adding a reservoir and a draining route to the network, can prevent 
the existence of a stable equilibrium state.
%To illustrate this, consider the following  idealized network 
%\begin{equation}
%\left.\begin{aligned}
%{\rm A}           & \rightarrow {\rm B} + {\rm B} & \quad k_1 & \,\,({\rm s}^{-1}) \\
%{\rm B} + {\rm B} & \rightarrow {\rm A}           & \quad k_2 & \,\,(\ccs) \\
%{\rm A}           & \rightarrow {\rm C}           & \quad k_3 & \,\,({\rm s}^{-1}) \\
%{\rm C}           & \rightarrow {\rm B} + {\rm B} & \quad k_4 & \,\,({\rm s}^{-1}) \\
%\end{aligned}
%\right.
%.
%\end{equation}
%If the abundance of A is a parameter and if we consider B out of equilibrium (to ensure 
%matter conservation), it is easy to show that the system has no fixed point with positive 
%abundances if $n({\rm A}) \geqslant n_{\rm B} / 2 \times [1 + k_3 / k_4]^{-1}$, where 
%$n_{\rm B} = 2n({\rm A}) + n({\rm B}) + 2n({\rm C})$ is the elementary density of B.
%When applied to our problem the previous example translates as follows:
For instance, if the abundance 
of \HH\ is too high, the steady state abundances of all other H-bearing species produced 
directly or indirectly via \HH\ cannot be small enough to guarantee the conservation of H 
atoms. In some cases, fixing the abundance of \HH\ can thus maintain the entire system out 
of equilibrium.

To identify such cases, we explored the convergence of the solver over a grid of 2.56 
million models. The results displayed in Fig. \ref{Fig-fail} reveal two regions in the 
4D parameter space.

The most pre-eminent region corresponds to high-density ($\dens \gtrsim 10^3$ \cc), shielded 
($A_V \gtrsim 1$) environments with moderate molecular fraction ($10^{-1} \lesssim 
f(\HH) \lesssim 1$) and kinetic temperature ($300\,{\rm K} \lesssim T_K \lesssim 10^3
\,K$). Testing those models without fixing the abundance of \HH\ shows similar results, 
which indicates that convergence failures are due to the Newton-Raphson algorithm and 
the topology of the Jacobian of system \ref{EqChemSys} rather than the absence of fixed 
point for the chemistry. When such cases occur, we therefore choose to guide the algorithm 
by computing the solution at high temperature and progressively decrease the temperature
towards the target value.

The least pre-eminent region in Fig. \ref{Fig-fail} corresponds to low-density ($\dens 
\lesssim 1$ \cc), illuminated ($A_V \lesssim 0.5$) environments with a high molecular 
fraction ($f(\HH) \gtrsim 0.95$) and kinetic temperature ($T_K \gtrsim 10^3\,K$). As 
predicted above, convergence failures happen here because the abundance of \HH\ is 
fixed and higher than a critical value. However, since such physical conditions are 
never reached in the simulation, we find that fixing the abundance of \HH\ and 
${\rm S}_{\rm H}$ never precludes the existence of an equilibrium state for the rest 
of the chemistry.

\subsubsection{Chemical timescales} \label{ap-timescales}

\begin{figure*}[!ht]
\begin{center}
\includegraphics[width=0.8\textwidth]{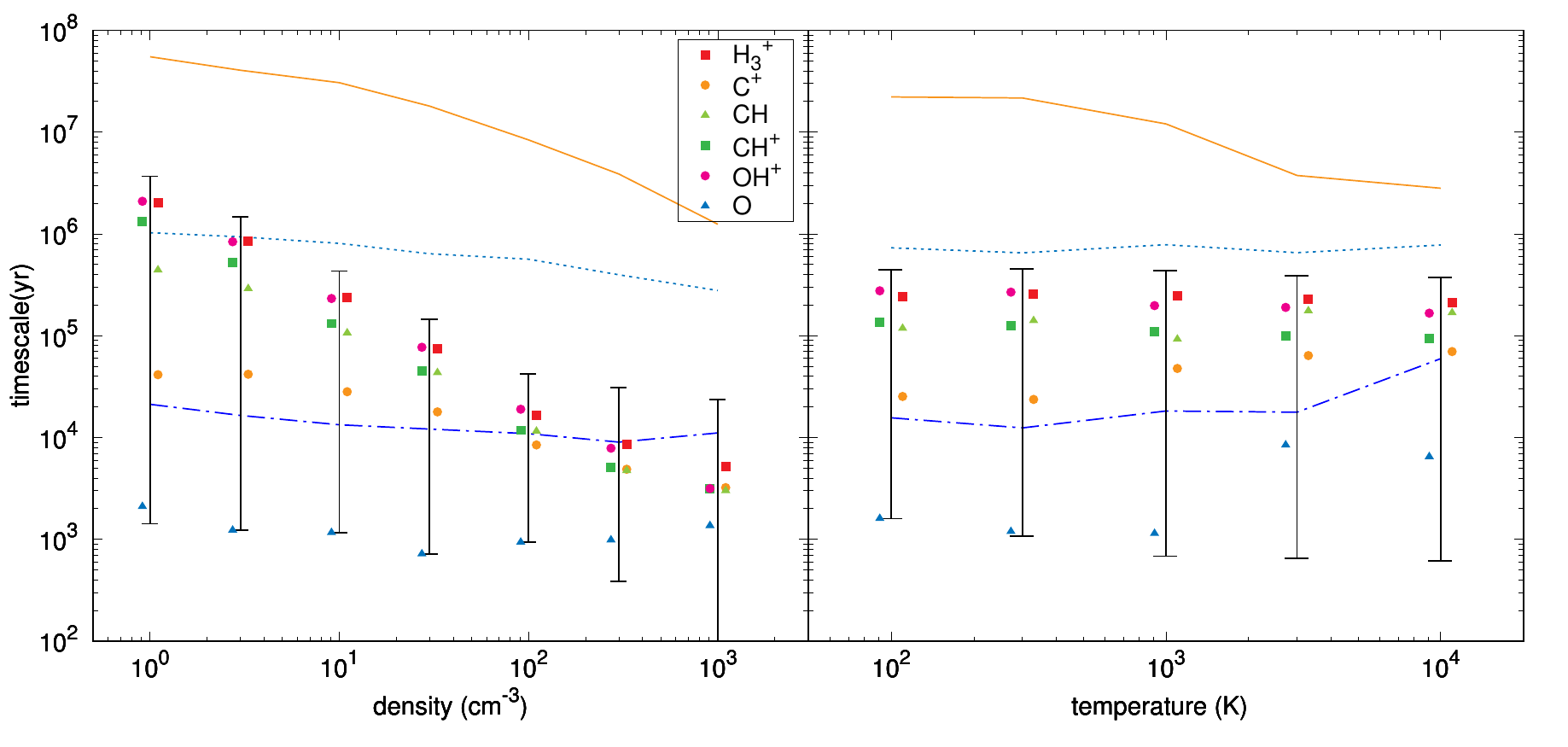}
%\includegraphics[width=18.0cm,angle=-0]{Figures/Grid_tscale_paper.pdf}
%\caption{Chemical timescale averaged over ten random initial conditions, two values of 
%$A_V$ (0.1 and 1), and four values of $f(\HH)$ ($10^{-3}$, $10^{-2}$, $10^{-1}$, and $1$, 
%for configuration b only, see main text) as functions of the density assuming a temperature 
%of 50 K (left panel) and as functions of the temperature assuming a density of 10 \cc\ 
%(right panel). The \HH\ timescales (colored curves) are computed in configuration (a) 
%(see main text) for $f_{\rm sh,\,\,\HH} = 10^{-4}$ (solid), $10^{-2}$ (dotted), and $1$
%(dotted dashed). The timescales of single molecules (colored points) and the range of 
%timescale of all other species (black segments) are computed in configuration (b) (see 
%main text).}
\caption{Timescales required by the species to reach their equilibrium abundances as a function of
the density of the gas for a temperature fixed at $50~\mathrm{K}$ (\emph{left panel}) and as a function of 
the temperature of the gas for a density of $10~\cc$ (\emph{right}). The \HH\ timescales (coloured 
curves) are computed in configuration (a) (see main text) for $f_{\rm sh, \HH} = 10^4$ 
(solid), $10^2$ (dotted), and $1$ (dot-dashed). Each curve results from an average of 
timescales computed for ten random initial conditions and two values of $A_V$ (0.1 and 1). 
The timescales of single molecules (coloured points) and the range of timescales of all 
other species (black segments) are computed in configuration (b) (see main text). These 
timescales are averaged over ten random initial conditions, two values of $A_V$ (0.1 and 1), 
and four values of $f(\mathrm{H_2})$ ($10^{-3}$, $10^{-2}$, $10^{-1}$, and $1$).}
\label{Fig-tscale}
\end{center}
\end{figure*}

Finally, fixing \HH\ presumes that the timescales required for each species to reach 
their equilibrium state in a medium at constant molecular fraction is smaller than that 
required for \HH. To test this hypothesis we have computed the equilibrium timescales 
$\tau({\rm X})$ of every species {\rm X} using the time-dependent integrator of the 
TDR code \citep{godard2014} set up to follow the chemical evolution of a fluid cell with 
constant density and temperature.  We define the timescale $\tau({\rm X})$ as the longest 
time for which the abundance of species X differs from its asymptotic value by more than 
10\%. With this definition, the integrator was run over grids of models of different 
density, temperature, and extinction in two different configurations: (a) considering all 
species as variables and adopting the \HH\ self-shielding factor $f_{\rm sh,\,\,\HH}$ as 
a parameter and (b) assuming a constant molecular fraction $f(\HH)$. All models were 
finally run with ten different initial conditions randomly drawn on a logarithmic scale
and rescaled to match the input elemental abundances (\citealt{glover2010} Appendix A).

The chemical timescales averaged over initial conditions, visual extinction, and \HH\
molecular fraction (for configuration b only) are displayed in Fig. \ref{Fig-tscale} as 
functions of the density and the temperature of the gas. Dispersion analysis shows that 
these equilibrium timescales are computed with an uncertainty of about a factor of 3. As 
expected, the equilibrium timescale of \HH\ derived in configuration (a) depends strongly 
on the self-shielding parameter, but only weakly on the density and the temperature of the 
gas. In contrast, the chemical timescales derived in configuration (b) are found to span 
a wide range of values (three orders of magnitude), to strongly depend on the gas density, 
and weakly depend on the gas temperature.

Figure \ref{Fig-tscale} shows that \HH\ is among the species with the longest evolution
times. The equilibrium timescale of \HH\ is found to be larger than that of any other 
species over a wide range of physical parameters, as long as $\dens \geqslant 3$ \cc\ 
or $f_{\rm sh,\,\,\HH} \leqslant 10^{-2}$. These wide conditions are fulfilled in 
more than 90 \% of the mass of the gas (\citetalias{valdivia2016}, Fig. 7), 
but also in more than 92 \% of the environments where CH$^+$ is produced 
(see Fig. \ref{CHp_dist_fig}). It therefore not only validates the existence of a chemical 
equilibrium in a medium with a fixed molecular fraction, but also justifies our approach 
to single out \HH\ in the chemistry.% Interestingly, it also indicates that the method, as 
%appropriate as it might be for H and \HH, won't be applicable to any other atoms or molecules.

It is, finally, important to note that these considerations justify  treating the chemistry 
at equilibrium only with respect to \HH, but not necessarily with respect to the gas dynamics. 
A more detailed comparison between the chemical timescales and the dynamical resolution of 
the simulation and the importance of transient events are discussed in  Sect. \ref{sect-timescale}.

\end{document}